\documentclass[lettersize,journal]{IEEEtran}
\usepackage{amsmath,amsfonts}
\usepackage{algorithmic}
\usepackage{array}
\usepackage[caption=false,font=normalsize]{subfig}
\usepackage{textcomp}
\usepackage{stfloats}
\usepackage{url}
\usepackage{verbatim}
\usepackage{graphicx}

\usepackage{enumitem}
\usepackage[ruled,vlined,linesnumbered]{algorithm2e}
\usepackage{booktabs}
\usepackage{cite}
\usepackage{caption}
\usepackage{stfloats}
\usepackage{makecell}
\usepackage{makecell,array}  
\usepackage{multirow,makecell}

\def\BibTeX{{\rm B\kern-.05em{\sc i\kern-.025em b}\kern-.08em
    T\kern-.1667em\lower.7ex\hbox{E}\kern-.125emX}}
\usepackage{balance}

\begin{document}
\title{Aidos: A Hybrid Optimization Algorithm for Beam Hopping Scheduling in NGSO Mega-Constellations}
\author{Lingkai~Zhao,
        ~Zhe~Chen,~\IEEEmembership{Member,~IEEE},%
        ~Kun~Qiu,~\IEEEmembership{Senior Member,~IEEE},%
        ~and~Yue~Gao,~\IEEEmembership{Fellow,~IEEE}%
\thanks{This work was sponsored by Natural Science Foundation of Shanghai under Project No. 25ZR1402021. \textit{(Corresponding author: Kun Qiu.)}
Lingkai Zhao, Zhe Chen, Kun Qiu and Yue Gao are with the School of Computer Science, Fudan University, Shanghai 200438, China, and also with the Institute of Space Internet, Fudan University, Shanghai 200438, China (e-mail: lkzhao24@m.fudan.edu.cn; zhechen@fudan.edu.cn; qkun@fudan.edu.cn; gao.yue@fudan.edu.cn).}   
}

\markboth{IEEE Transactions on Mobile Computing}%
{Aidos: A Hybrid Optimization Algorithm For Beam Hopping Scheduling in NGSO Mega-Constellations}

\maketitle
\begin{abstract}
With the rapid proliferation of non-geostationary orbit (NGSO) mega-constellations, beam hopping (BH) has become indispensable for resource scheduling in multi-satellite, multi-coverage scenarios. By dynamically adjusting spot beam power and pointing within each time slot, BH enables highly efficient spectrum utilization. A principal engineering challenge is the real-time generation of beam hopping time plans (BHTP). 
Traditional algorithms, such as the round-robin strategy, distribute beams evenly across all service cells in a round-robin fashion. 
However, real traffic follows a long-tail distribution; the most active 10\% of hotspot cells generate more than 50\% of the aggregate demand, making uniform allocation inadequate.
To address this issue, existing frameworks adopt a genetic algorithm (GA), whose throughput is approximately 80.7\% higher than the traditional baseline.
Operational satellite footprints encompass more than 1,000 service cells. The GA requires 67.8 s to generate a BHTP for 1,127 cells. With a 550 km LEO satellite providing only a 300 s visibility window, multiple online recomputations are impractical.
State-of-the-art algorithms, such as multi-agent deep reinforcement learning (MADRL), fail to converge once the cell count exceeds 200.
To overcome these challenges, we propose a novel BH scheduling algorithm Aidos.
The algorithm integrates traffic-aware random-key encoding into a multi-objective metaheuristic search, and then applies a sliding-window Beta resampling strategy during adaptive distribution evolution, to improve both the search efficiency and the solution quality of the BHTP. 
Experiments demonstrate that Aidos improves throughput by 79.2\% and reduces latency by 99.45\%. Its average computation time is 9.3 s, enabling online replanning within a 300 s satellite overpass window.
\end{abstract}


\IEEEdisplaynontitleabstractindextext
\IEEEpeerreviewmaketitle
\section{Introduction}
\IEEEPARstart{W}{ith} the rapid growth of non-geostationary orbit (NGSO) mega-constellations such as Starlink, OneWeb, and Kuiper, on-demand resource scheduling and interference mitigation under high mobility and multi-coverage conditions have become essential\cite{kodheli2020satellite}\cite{al2022survey}. To address these challenges, standardization bodies including 3GPP Releases 17/18/19 and DVB-S2X have advanced phased-array-based beam hopping (BH) technology \cite{3gpp2020study}\cite{DVB-S2X2020}. At the timeslot scale, BH reconfigures the pointing direction and transmit power of multiple spot beams in a time-division manner, focusing on hotspot cells as needed and reducing idle capacity in low-load regions. With appropriate dwell time and scheduling, BH can significantly increase system capacity \cite{rohde2018beam}.

A critical challenge in deploying BH technology is the real-time generation of an optimal beam hopping time plan (BHTP)\cite{DVB-S2X2020}. 
In multi-satellite cooperative scenarios, the goal of BH scheduling is to maximize throughput and minimize latency while meeting engineering constraints such as interference limits and revisit time.
The resulting optimization problem is a non-convex nonlinear mixed-integer programme \cite{cocco2017radio}, proven to be NP-hard \cite{aravanis2015power}.
The decision-space size grows approximately with the product of cells, beams, time slots, and satellites.
For example, a single Starlink satellite can cover about 1,000 cells \cite{spacex2017filing}.
Multi-satellite cooperation increases the problem scale to $10^{3}$-$10^{4}$, which substantially raises computational cost and solution difficulty.

The core task of a BHTP is to compute an onboard resource allocation scheme that includes carrier frequency, transmit power, beam index and time slot, in order to satisfy the traffic demands of terrestrial cells.
In practical implementations, the most straightforward approach is the round robin strategy \cite{anzalchi2010beam}. 
Uniform allocation cannot track spatial demand heterogeneity. 
Hotspots receive insufficient resources and low-load cells waste capacity, which reduces aggregate throughput.
To enhance system performance, prior studies have explored meta-heuristic search algorithms such as the multi objective genetic algorithm (MOGA-BH) \cite{deng2024satellites} and the artificial bee colony algorithm \cite{wei2022dynamic}. These methods, however, typically optimize only a single objective and report solely offline results, without fully accounting for real-time applicability. Experimental results show that MOGA-BH improved throughput by approximately 80.7\% over the conventional baseline in a 1,127-cells scenario. However, generating a single BHTP still required 67.8 s, which is impractical for operational systems.

The state-of-the-art method adopts a multi-agent deep reinforcement learning (MADRL) approach, in which each satellite employs a policy trained with the QMIX algorithm to collaboratively generate real-time BH decisions\cite{lin2024satellite}\cite{kim2025dqn}. This MADRL framework achieves millisecond-level inference latency and schedules beams effectively, but it has been validated only in a scenario with 76 ground cells. In large-scale LEO constellations, however, a single satellite may cover thousands of cells and is equipped with 48 independent downlink beams \cite{spacex2017filing}. As the numbers of cells and beams surge, the action space grows exponentially, triggering the curse of dimensionality. Experimental results show that when the number of cells exceeds 200, gradient vanishing occurs and the model fails to learn effective policies. Consequently, such MADRL methods are unsuitable for large-scale LEO satellite scenarios.

To overcome the above limitations and generate BHTP under realistic NGSO constellation coverage, we propose Aidos. 
The algorithm takes real traffic as input and outputs an executable BHTP. 
Motivated by the high-dimensional and discrete nature of BHTP, we design a traffic-aware random-key encoding that embeds discrete traffic demands into a continuous key space. 
Building on this representation, we construct a multi-objective metaheuristic search that jointly optimizes throughput, delay, and fairness.
We further introduce a sliding-window Beta resampling strategy to accelerate convergence and improve solution quality.
Our evaluation indicates that Aidos outperforms MOGA-BH and MADRL-BH by 79.2\% and 247.2\% in throughput, respectively, while reducing latency by 99.45\%. Its average computation time is 9.3 s, enabling online replanning within a 300 s satellite overpass window.

Briefly speaking, this paper makes the following contributions:
\begin{itemize}
    \item[\indent $\bullet$] We propose Aidos, an improved metaheuristic search algorithm. The algorithm can be used for computing BHTP in large-scale scenarios.

    \item[\indent $\bullet$] We design a traffic-aware random-key encoding to handle discrete traffic inputs and build a multi-objective metaheuristic search to compute the optimal BHTP. In addition, we introduce sliding-window Beta resampling and multi-scale perturbation to accelerate convergence and improve the quality of the optimal solution.

    \item[\indent $\bullet$] The proposed algorithm is evaluated from four perspectives: overall performance, convergence behavior, scalability and computation time. Comparative results against multiple baseline algorithms confirm that Aidos achieves superior throughput and delay performance. Moreover, the framework supports online computation in large-scale scenarios.
    
\end{itemize}

The rest of the paper is organized as follows: Section II introduces the background of beam hopping scheduling. Section III demonstrates the overview design of Aidos. Section IV describes the detailed design of the core modules in Aidos. Section V evaluates Aidos and analyzes the evaluation results. Finally, Section VI concludes.

\section{Background}
\subsection{Beam Hopping (BH)}
\begin{figure}[!h]
\centering  
\includegraphics[width=0.95\linewidth, clip, 
                trim=0.9cm 0.9cm 0.9cm 0.9cm]{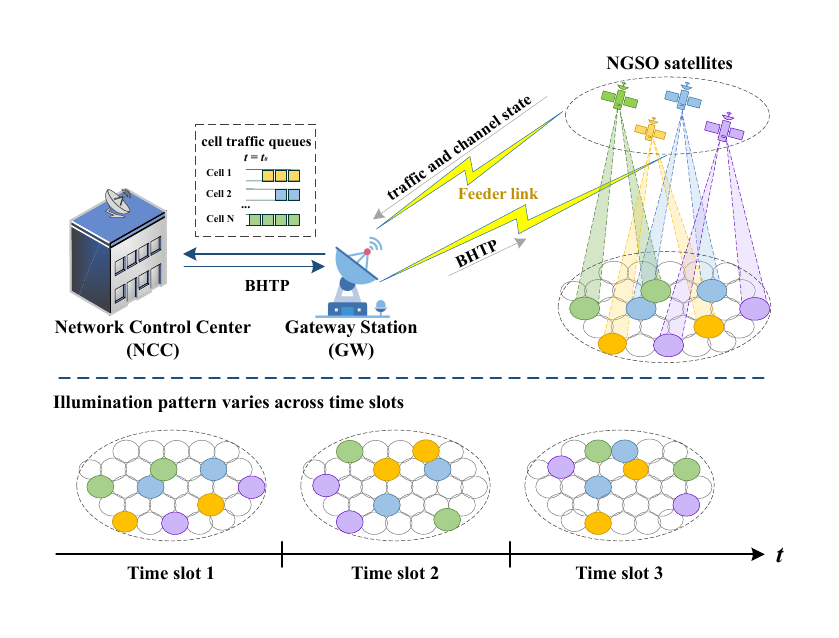} 
\caption{Three layers of BH systems: ground planning layer, air-interface signalling layer and on-board execution layer.}
\label{fig:BH system}
\vspace{-0.3cm}
\end{figure}

Beam hopping (BH) is a time-division satellite resource-scheduling technique.
Using phased-array antennas, a satellite steers a limited set of high gain spot beams to different ground cells in successive time slots to match the spatial traffic distribution. Compared with fixed multibeam systems, BH concentrates spectrum and amplifier power on hotspot areas while reducing idle capacity in lightly loaded regions, thereby improving instantaneous throughput and spectral efficiency. As shown in Fig.~\ref{fig:BH system}, 
BH systems usually comprise three layers\cite{DVB-S2X2020}:
\begin{itemize}
    \item[\indent $\bullet$] Ground planning layer: The satellite collects real-time traffic and channel information and downlinks it to the gateway, where a BHTP is computed. The BHTP specifies the service order and dwell time of each beam.

    \item[\indent $\bullet$] Air-interface signalling layer: The generated BHTP is encapsulated in the superframe header and other control fields defined in Annex E of DVB-S2X, and then forwarded to the satellite over the feeder link.

    \item[\indent $\bullet$] On-board execution layer: A digital beam-former dynamically adjusts the antenna, switching the beam pointing direction and transmit power in real time according to the BHTP.
\end{itemize}

\subsection{Beam Hopping Time Plan (BHTP)}
\begin{figure}[!h]
\centering  
\includegraphics[width=0.8\linewidth, clip, 
                trim=0.7cm 0.7cm 0.7cm 0.7cm]{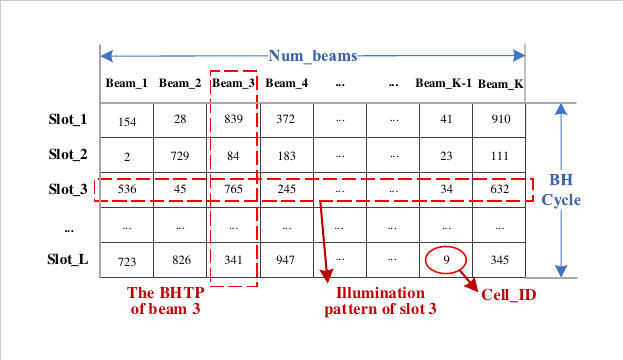} 
\caption{A complete BHTP for $K_{\max}$ beams across $L$ time slots. Each row corresponds to a time slot, each column to a beam\_ID, and each matrix element to a cell\_ID.}
\label{fig:BHTP}
\vspace{-0.4cm}
\end{figure}

In a pre-scheduled DVB-S2X BH system, several beam hopping traffic channels (BHTC) run in parallel and repeat periodically to serve different cell clusters. Each BHTC illuminates the cells in its cluster according to a beam hopping time plan (BHTP). A BHTP specifies, for every time slot in one complete hopping period, two items: 1) the cell to be illuminated and 2) its dwell time \cite{DVB-S2X2020}. For simplicity, we set the dwell time equal to the slot length \cite{duyckdemonstrating}. Thus, the BHTP fully describes the slot-by-slot illumination sequence. After a satellite receives a BHTP from the gateway, it executes the table in a loop until a new BHTP arrives. Fig.~\ref{fig:BHTP} shows the detailed structure of a BHTP.

\subsection{Slot and Cycle Configuration}
The physical layer follows the DVB-S2X Annex E superframe configuration, and the system operates in a burst-mode downlink.
Based on Starlink FCC reports, we set the single-beam net bit rate to 200 Mbps.
We adopt 16-APSK with a 3/4 code rate as the baseline MODCOD.
Annex E defines a fixed superframe length of \(L_{\mathrm{SF}} = 612{,}540\) symbols.
This value implies a physical lower bound of about 10 ms for the BH dwell time. This bound preserves full alignment of the synchronization fields.

\begin{figure}[!h]
\centering  
\includegraphics[width=0.9\linewidth, clip, trim=0cm 1cm 1cm 1cm]{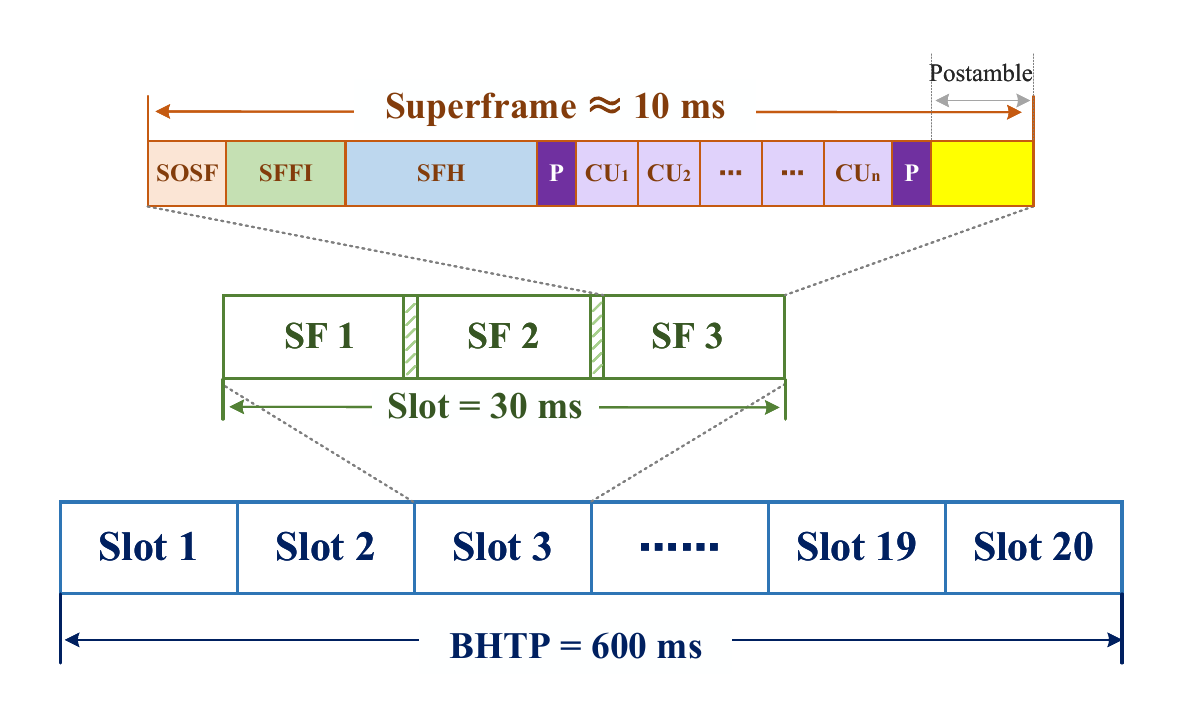} 
\caption{Illustration of the hierarchical relationship among the BH period, time slot, and superframe structure.}
\label{fig:sf}
\vspace{-0.2cm}
\end{figure}

We set the BH slot to 30 ms. The slot aggregates three superframes and reserves about 2 ms as a guard interval to absorb clock drift and reduce beam-switching overhead.
Within the 30 ms slot, the system can switch among multiple MODCODs, such as 16-APSK 3/4 and 8-PSK 5/6, which improves link adaptation.
To meet the cell revisit time constraint, we set a BH cycle of 20 slots and a total duration of 600 ms, as shown in Fig.~\ref{fig:sf}.
A satellite at 550 km altitude has a visibility window of about 300 s, which is much longer than this cycle. 
The configuration provides sufficient spatiotemporal resolution for dynamic traffic.

\subsection{Particle Swarm Optimization (PSO)}
The proposed Aidos is inspired by particle swarm optimization (PSO).
PSO is a population-based metaheuristic motivated by swarm intelligence. The algorithm maintains a swarm of particles. Each particle has a position and a velocity, which represent a candidate solution and its update direction. A particle moves using two sources of information: its personal best position \(p_{\text{best}}\) and the best position observed by the swarm \(g_{\text{best}}\).
The main procedure is as follows. First, evaluate the fitness of $N_p$ particles.
Next, each generation iterates through the loop of \textit{evaluation $\rightarrow$ update of \(p_{\text{best}}\) and the Pareto archive $\rightarrow$ individualized selection of \(g_{\text{best}}\) $\rightarrow$ particle perturbation}. 
Once a preset iteration limit is reached, select the best Pareto solution from the elite archive as the output.

\subsection{Previous Frameworks and Issues}
Various standards organizations have supported BH technology from different perspectives. DVB-S2X Annex E defines a beam hopping oriented superframe (SF) at the air-interface layer and specifies how the BHTP is carried in the control fields. ITU-R provides reference radiation patterns for NGSO multi-beam antennas operating in the Ku bands. 3GPP Releases 17/18/19 enhance the beam management procedures in NR-NTN to accommodate the rapid beam switching required on the satellite side.

Early practical systems adopted round-robin scheduling\cite{feltrin2016eutelsat} and random scheduling\cite{freedman2017beam}.
These approaches are easy to implement but yield poor throughput. Greedy heuristics \cite{tian2019efficient} offer fast computation as well. However, they ignore nonlinear coupling and lack global information, so they cannot reach the maximum throughput in an NP-hard search space.
State-of-the-art studies employ metaheuristic search with GA \cite{deng2024satellites} and MADRL. 
GA has high computational complexity and cannot support online computation in large-scale scenarios. MADRL also fails to converge at scale. 
Value-based methods, such as Q-learning and DQN, require maximization over a discrete action set.
In large-scale scenarios, the per-slot action space grows to $\binom{N_c}{B}$, making such methods difficult to implement \cite{lin2024satellite}     
\cite{kim2025dqn}           
\cite{chen2025distributed}  
\cite{gong2026distributed}  
\cite{xu2025service}        
\cite{ouyang2025dependency}.   
Actor–critic methods such as MAPPO rely on a centralized critic, whose growing global state and joint action spaces hinder efficient and stable training
\cite{huang2025hybrid}      
\cite{lai2025dynamic}.       
To address these issues, we propose Aidos to compute the BHTP efficiently in large-scale settings.

\begin{figure}[t]
\centering  
\includegraphics[width=1\linewidth, clip, trim=0.95cm 0.95cm 0.85cm 0.85cm]{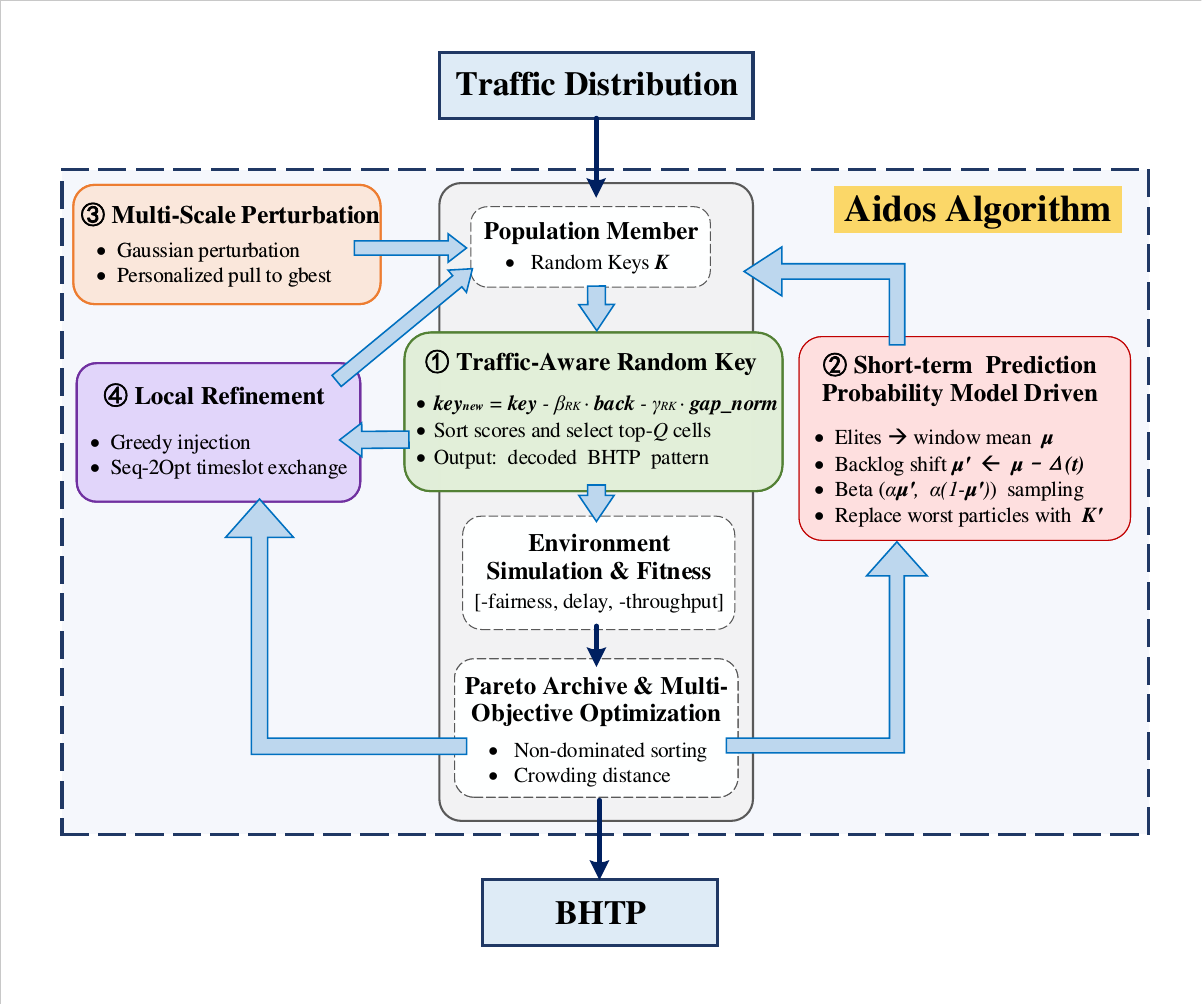} 
\caption{The overall architecture of Aidos}
\label{fig:Aidos}
\vspace{-0.3cm}
\end{figure}

\section{Design Overview }
To compute the BHTP in large-scale scenarios, we propose a hybrid optimization algorithm Aidos. 
The algorithm employs a multi-objective, population-based search that iteratively improves the BHTP.
In each generation, it evaluates the fitness of the population, updates personal bests and the Pareto archive, and then assigns a personalized reference solution to guide the next update.
The outer-loop search is configured by the swarm size $N_p$, the iteration budget $N_{\mathrm{iter}}$, and the update coefficients $(w_{\mathrm{in}},c_1,c_2)$ in Table~II.
To enhance search capability, Aidos integrates four cooperative modules for encoding, probability guidance, local acceleration, and step-size adaptation. 
These modules enable fast convergence to high-quality BHTP solutions in large-scale scenarios.
The overall structure of the algorithm is illustrated in Fig.~\ref{fig:Aidos}.

\subsection{Traffic-Aware Random Key Encoding Module (TARK)}
As shown in Fig.~\ref{fig:BHTP}, the BHTP can be modeled as a high-dimensional discrete matrix. Each entry stores a cell\_ID.
Discrete decisions do not fit population-based search. Aidos therefore introduces a TARK module. 
For each slot–cell pair, the module generates a key in $(0,1)$ and forms a continuous key matrix with shape [num\_slots, num\_cells].
This matrix encodes the service-priority order of all cells.
During decoding, the algorithm computes a composite priority from the key value, the current backlog weight, and the revisit interval.
The system then selects covered cells for each time slot according to this priority and obtains a BHTP that satisfies the constraints.
The method preserves the solution-space structure and embeds hotspot preference and fairness in the continuous domain, which makes the search process smoother.

\subsection{Short-term Prediction Probability Model Driven Resampling Module (STP-PMD)}
Every $T_{\mathrm{PMD}}$ generations, where $T_{\mathrm{PMD}}$ is the trigger interval of STP-PMD, Aidos activates this module after fitness evaluation and Pareto-archive update to improve subsequent BHTP iterations. 
The module selects the top $\rho=20\%$ elite random-key matrices from the
Pareto archive to estimate slot-cell selection patterns.
These elites carry useful spatiotemporal priors about effective service of slot–cell pairs.
By computing sliding-window means of the keys, the module estimates slot–cell occupancy probabilities empirically.
It then shifts the mean using short-term traffic predictions and backlog traffic.
Using the corrected occupancy probability $\mu^{\prime}$, we build a probabilistic model $\mathrm{Beta}(\alpha \mu^{\prime},\mathrm{~}\alpha(1-\mu^{\prime}))$ with prior information.
Here, $\mathrm{Beta}(a,b)$ denotes the Beta distribution on $(0,1)$ with shape parameters $a>0$ and $b>0$ (details in Section IV.C).
We sample new random-key matrices from this model and replace the worst-fitness solutions.
By estimating slot–cell occupancy probabilities and periodically injecting layouts that match hotspot and temporal statistics, the module accelerates multi-objective convergence.

\subsection{Local Refinement Module (LR)}

The LR module is triggered every $T_{\mathrm{LR}}$ generations to rapidly advance the Pareto front.
The module first selects individuals with high crowding and poor fitness. It then applies two direct operations to their BHTP: 1) greedy injection: overwrite selected beam columns with the set of cells that have the largest current backlog to relieve queue pressure; 
2) Seq-2Opt: swap two entire time-slot rows within \(|t_{1}-t_{2}| \leq 3\) and accept the swap if the estimated throughput increases.
After the column overwrite and the row swap, the BHTP obtains a local improvement.

\subsection{Multi-Scale Perturbation Module (MSP)}
The MSP module operates on the BHTP random-key matrix at the end of each generation and preserves exploration through multi-scale perturbations.
First, it adds Gaussian noise to the full matrix to generate diversified slot--cell priority patterns.
Second, with a fixed probability, it moves the perturbed matrix toward an individualized guiding solution to avoid excessive drift.
Finally, it updates the private step size $\sigma_i$ of particle $i$ by the 1/5 success rule, where $\sigma_i$ controls the magnitude of subsequent Gaussian perturbations on individual key values.
The procedure preserves diversity and improves convergence efficiency, thereby yielding high-quality BHTP solutions.
\begin{table}[t]
  \centering
  \caption{NOTATION AND DESCRIPTION}
  \label{tab:notation}
  \begin{tabular}{|c|p{0.7\linewidth}|}
    \hline
    Notation & Definition \\ \hline
    $\mathcal{N}$     & Number of satellites \\ 
    $\mathcal{M}$     & Number of cells \\ 
    $K_{\max}$        & Maximum active beams per satellite \\
    $Q$     & Total number of schedulable beams per slot \\ 
    $\mathcal{L}$     & Number of time slots \\ 
    $\mathbf{B}$      & Global BHTP matrix \\ 
    $T_{\mathrm{slot}}$   & Duration of one time slot \\ 
    $T_{\mathrm{cycle}}$  & Duration of one BHTP cycle \\ 
    $D_{m,i}$       & Traffic demand of cell $m$ at slot $i$ \\
    $D_{m,0}$       & Initial traffic demand of cell $m$ at the start of a cycle \\
    $C_{m,i}$       & Channel capacity of cell $m$ at slot $i$ \\
    $A_{m,i}$         & Rate relaxation gain of cell $m$ during slot $i$ \\
    $L_{\mathrm{pkt}}$& Fixed packet length \\
    $\lambda_{m,i}^{\mathrm{tot}}$ & Total packet arrival rate of cell $m$ at slot $i$ \\
$\lambda_{m,i}^{\mathrm{rt}}$  & Real-time packet arrival rate of cell $m$ at slot $i$ \\
    $\mu_{m,i}$       & Service rate of cell $m$ at slot $i$ \\
    $b_{i,q}$         & Cell assigned to global beam $q$ at slot $i$ of the BHTP \\
    $F_1,F_2,F_3$     & Fairness, delay, and throughput objective \\ \hline
  \end{tabular}
\end{table}

\section{Algorithm Design}
In this section, we introduce the system model of the Aidos and the design details of each module.

\subsection{Problem Formulation}

This study focuses on centralized BH scheduling for the downlink of an NGSO constellation. Let the minimum receive elevation angle at the ground control center be~$\theta$. 
At any slot, the number of visible satellites is~$N$, indexed by $n=1,2,\ldots,N$. The control center allocates beams to these $N$ satellites in a unified manner. The Earth’s surface is discretized into hexagonal cells using the H3 geospatial indexing library at resolution level 5, yielding a cell area that closely matches the coverage footprint of a Starlink spot beam. Superimposing the instantaneous fields of view of all visible satellites yields $M$ ground cells, indexed by $m=1,2,\ldots,M$. Each satellite can illuminate at most $K_{\max}$ spot beams simultaneously. Therefore, the total number of schedulable beams in each slot is $Q = N K_{\max}$.

The satellite--ground communication model adopted in this study follows the generic framework in 
\cite{lin2024satellite}
\cite{kim2025dqn}.
The BHTP is the core decision variable in Aidos and is defined as
\begin{equation}
\mathbf{B}=\begin{bmatrix}
b_{i,q}
\end{bmatrix}\in\{0,1,2,\ldots,M\}^{L\times Q}
\end{equation}
where $i=1,\ldots,L$ and $q=1,\ldots,Q$ denote the slot and global beam indices, respectively. 
The $Q$ columns are grouped by satellite, with columns $(n-1)K_{\max}+1$ to $nK_{\max}$ corresponding to satellite $n$. 
Each element $b_{i,q}$ denotes the assignment of beam $q$ in slot $i$, where $b_{i,q}=m$ means serving cell $m$ and $b_{i,q}=0$ means idle. 
Moreover, $L=T_{\text{cycle}}/T_{\text{slot}}$.

The optimization objectives considered in this work comprise fairness, average real-time packet delay, and throughput. Once the BHTP has been determined, these three metrics are evaluated at the end of each scheduling cycle. The fairness metric is defined as follows:
\begin{equation}F_1(\mathbf{B})=\sum_{m=1}^M\frac{\sum_{i=1}^LC_{m,i}(\mathbf{B})T_{\mathrm{slot}}}{D_{m,0}+\sum_{i=1}^LA_{m,i}}\end{equation} where \( D_{m,0} \) denotes the unserved traffic demand of cell \( m \) at the start of the cycle; \( C_{m,i}(\mathbf{B}) \cdot T_{\text{slot}} \) is the amount of data served to cell \( m \) during slot \( i \); and \( A_{m,i} \) represents the rate relaxation gain at cell \( m \) during slot \( i \), assumed to follow a Poisson distribution.

Consistent with prior research, the BH service process is commonly modeled as an M/M/1 queuing system~\cite{zheng2024traffic}, where the service rate of a downlink beam is given by
\begin{equation}
\mu_{m,i}(\mathbf{B})=\frac{C_{m,i}(\mathbf{B})}{L_{\mathrm{pkt}}}
\end{equation}
where \(L_{\mathrm{pkt}}\) denotes the fixed packet length. 
For cell $m$, the total packet arrival rate at slot $i$ is denoted by $\lambda_{m,i}^{\mathrm{tot}}$.
When \(\lambda_{m,i}^{\mathrm{tot}} < \mu_{m,i}(\mathbf{B})\), the average queuing delay is given by
\begin{equation}
W_{m,i}(\mathbf{B})=\frac{1}{\mu_{m,i}(B)-\lambda_{m,i}^{\mathrm{tot}}}
\end{equation}

Following the mixed-traffic profile reported by the NGMN Alliance~\cite{NGMN_RAN_Eval_Methodology_2008,NGMN_CRAN_Critical_Technologies_2015}, we assume that real-time traffic accounts for 30\% of the total demand. 
For real-time service with a delay deadline $\tau_{\max}$, packets whose queuing delay exceeds $\tau_{\max}$ are treated as dropped. 
Dropped packets are assigned a fixed penalty delay $\kappa \tau_{\max}$ in the objective, where $\kappa>1$ is a penalty factor and $\kappa=10$ in this paper. The penalized delay is defined as
\begin{equation}
\widetilde{W}_{m,i}(\mathbf{B})=
\begin{cases}
W_{m,i}(\mathbf{B}), & W_{m,i}(\mathbf{B}) < \tau_{\max},\\
\kappa \tau_{\max}, & W_{m,i}(\mathbf{B}) \ge \tau_{\max},
\end{cases}
\end{equation}

The average delay objective of real-time packets is then given by
\begin{equation}
F_2(\mathbf{B})=
\frac{\sum_{m=1}^{M}\sum_{i=1}^{L}\lambda_{m,i}^{\mathrm{rt}}\widetilde{W}_{m,i}(\mathbf{B})}
{\sum_{m=1}^{M}\sum_{i=1}^{L}\lambda_{m,i}^{\mathrm{rt}}}
\end{equation}
where the average arrival rate of real-time traffic is \(\lambda_{m,i}^{\mathrm{rt}} = 0.3\,\lambda_{m,i}^{\mathrm{tot}}\).

The total system throughput within a BH cycle is given by
\begin{equation}
F_3(\mathbf{B})=\sum_{m=1}^{M}\sum_{i=1}^{L}C_{m,i}(\mathbf{B})T_{\mathrm{slot}}
\end{equation}

In summary, the following multi-objective optimization problem is formulated in this study:
\begin{equation}
\mathcal{P}_1:\min_{\mathbf{B}}
\quad \left[-F_1(\mathbf{B}),\, F_2(\mathbf{B}),\, -F_3(\mathbf{B})\right]
\end{equation}
\begin{align}
\text{C}_1:~ & \sum_{q=1}^{Q}\mathbf{1}_{\{b_{i,q}=m\}} \le 1,\quad \forall m,i,\\
\text{C}_2:~ & \sum_{q=(n-1)K_{\max}+1}^{nK_{\max}}
\sum_{m=1}^{M}\mathbf{1}_{\{b_{i,q}=m\}} \le K_{\max},
\quad \forall n,i.
\end{align}
where $F_1$, $F_2$, and $F_3$ are defined in (2), (6), and (7), respectively. Aidos optimizes this three-dimensional objective vector in a Pareto sense. Constraints C1--C2 ensure scheduling feasibility. Specifically, C1 enforces that each cell is served by at most one global beam in each slot. 
C2 is a structural feasibility constraint consistent with the global BHTP representation, ensuring that the beam assignments associated with satellite $n$ remain within its dedicated $K_{\max}$ beam-column block.


\begin{algorithm}[t]
  \caption{Traffic-aware Random-Key \,Decoding}
  \label{alg:RK-d}
  \KwIn{%
    key matrix $K_{L\times M}$; 
    number of slots $L$;
    number of cells $M$;
    number of beams $Q$;
    backlog weights $back$;
    $\Delta t$ counters $gap_{1..M}$; $\beta_{\mathrm{RK}},\gamma_{\mathrm{RK}}$; 
  }
  \KwOut{%
    illumination pattern $Pattern_{L\times Q}$; updated $gap$
  }
  \BlankLine
  \For{$t \gets 1$ \KwTo $L$}{%
      $gap\_norm \leftarrow
        \dfrac{gap-\min(gap)}{\max(gap)-\min(gap)+\varepsilon}$\;
      $score \leftarrow K[t,:]\;-\;\beta_{\mathrm{RK}}\,\cdot \mathbf{back}\;-\;\gamma_{\mathrm{RK}}\,\cdot \mathbf{gap\_norm}$\;
      $order \leftarrow \textbf{argsort}(score)$\;
      $sel \leftarrow order[1\ldots Q]$\;
      $Pattern[t,:] \leftarrow sel$\;
      $gap \leftarrow gap + 1;\quad gap[sel]\leftarrow 0$\;
  }

\end{algorithm}

\begin{algorithm}[t]
  \caption{Traffic-aware Random-Key \,Encoding}
  \label{alg:RK-e}
  \KwIn{%
    illumination pattern $Pattern_{L\times Q}$; 
    number of slots $L$;
    number of cells $M$;
    number of beams $Q$;}
  \KwOut{%
    encoded key matrix $K'_{L\times M}$
  }
  \BlankLine
  $K' \leftarrow \mathbf{0}_{L\times M}$\;
  \For{$t \gets 1$ \KwTo $L$}{%
      $ranks \leftarrow \textbf{rank\_ascending}(Pattern[t,:]) / M$\;
      $K'[t,Pattern[t,:]] \leftarrow ranks$\;
      $mask \leftarrow \{1,\ldots,M\} \setminus Pattern[t,:]$\;
      $K'[t,mask] \leftarrow 0.8 + 0.2 \cdot \textbf{UniformRand}(|mask|)$\;
  }
\end{algorithm}

\subsection{Traffic-Aware Random Key Encoding Module (TARK)}

Solving the BHTP is a high-dimensional discrete combinatorial optimization problem.
Directly updating particle velocity and position in the discrete space does not yield smooth improvement and often causes premature convergence.
Therefore, we design a traffic-aware random-key encoding that maps discrete assignments to a continuous domain. This representation improves exploration efficiency.

Specifically, each population member holds a random-key matrix $K$ with shape [num\_slots, num\_cells]. Each entry lies in the interval $(0,1)$. 
The value $key_{t,c}$ denotes the priority key for slot $t$ and cell $c$; a smaller value indicates a higher priority.
Optimization proceeds in the continuous key space and the discrete beam assignment is generated during decoding.

During decoding, we execute three steps for each time slot:

\textit{1)}  
Update the row vector of key values using backlog vector $\mathbf{back}$ and the unserved-interval vector $\mathbf{gap\_norm}$:
\begin{equation}
\mathbf{key}^{\mathrm{new}}_{t}
=
\mathbf{key}_{t}
-\beta_{\mathrm{RK}}\,\mathbf{back}
-\gamma_{\mathrm{RK}}\,\mathbf{gap\_norm}
\end{equation}
where $\beta_{\mathrm{RK}}$ and $\gamma_{\mathrm{RK}}$ are the weight coefficients of the random key module.

\textit{2)}  
Sort the row by the updated keys and select beams in ascending order to form the slot’s illumination pattern.

\textit{3)} 
Update $\mathbf{gap\_norm}$ and proceeds to the next slot.

After selecting beams for all time slots, we obtain the final BHTP.
The module embeds fairness based on backlog and waiting time directly in the decoding logic.
This design balances service frequency between hotspots and low-load regions in a dynamic manner.

\begin{figure*}[!t]
\centering  
\includegraphics[width=0.9\linewidth, clip, 
trim=0.6cm 0.7cm 0.8cm 0.8cm]{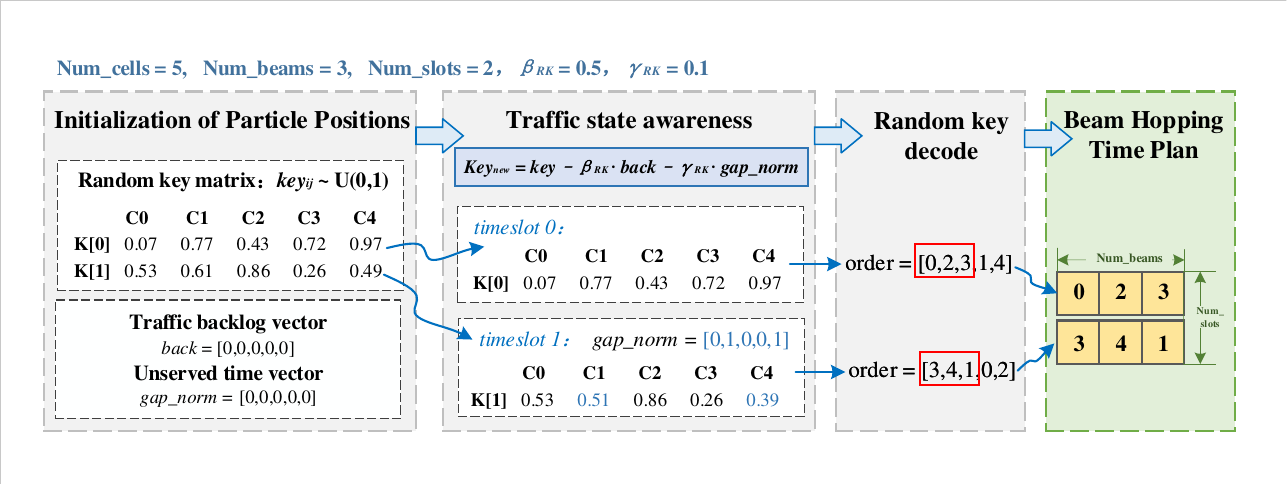} 
\caption{A numerical example of TARK module. Each particle initializes a random 2×5 key matrix representing five cells over two time slots. At time slot 0, the backlog is zero, so the original keys are retained. After sorting, the top three cells $C_0$, $C_2$, and $C_3$ are selected, and $\mathbf{gap\_norm}$ is updated to $[0,1,0,0,1]$. At time slot 1, only unserved cells $C_1$, $C_4$ have their key values reduced by 0.1 to increase their selection priority. This process yields the final BHTP.}
\label{fig:RK}
\vspace{-0.3cm}
\end{figure*}

Algorithm~\ref{alg:RK-d} presents the pseudocode of the TARK decoding. Line 2 computes the information required for decoding. It iterates over time slots and normalizes the revisit interval to obtain $\mathbf{gap\_norm}$.
Lines 3–5 perform priority evaluation and candidate selection.
Lines 6–7 update the state.

The time complexity of the decoding algorithm is summarized as follows:
line 2-3 require $O(M)$ time; at line 4, the argsort call costs $O(MlogM)$; line 5-7 contribute $O(M+Q)$. Since the loop executes for $L$ slots, the total complexity is $O(LMlogM)$.

Algorithm~\ref{alg:RK-e} presents the pseudocode of the TARK encoding.
Line 1 initializes the encoded key matrix $K'$ with zeros. 
Lines 2–4 assign very small keys to the cells selected in the current slot, in ascending rank order, and write them back to their positions.
Lines 5–6 assign large random keys in $[0.8,1.0)$ to the unselected cells.

The time complexity of the encoding procedure is as follows:
line 1 takes $O(LM)$; lines 2–4 together incur $O(QlogQ+Q)$; line 5-6 contribute $O(M)$. 
Over all $L$ slots, the total complexity is
$O(L(QlogQ+M))$.
Since $M\gg Q$ in practice, the dominant term is $O(LM)$.

\textit{Example Analysis:} For example, as shown in Fig.~\ref{fig:RK}, a random 2×5 key matrix is first generated for each particle, representing the continuous encoding of five cells across two time slots. At time slot $t$, the key values are updated based on $\mathbf{back}$ and $\mathbf{gap\_norm}$. In time slot 0, since the backlog is zero, the original key values are retained. After sorting the key values in ascending order, the top three cells $C_{0}, C_{2}$ and $C_{3}$ are selected for illumination, and the $\mathbf{gap\_norm}$ is updated to $[0, 1, 0, 0, 1]$. Upon entering time slot 1, the key values are updated again. It can be observed that only the cells $C_{1}$ and $C_{4}$, which were not served in the previous time slot, have their key values decreased by 0.1 to increase their relative priority in the current time slot's sorting. The final result is then used to compute the BHTP.

\subsection{Short-term Prediction Probability Model Driven Resampling Module (STP-PMD)}

\begin{figure*}[!t]
\centering  
\includegraphics[width=0.9\linewidth, clip, 
trim=1cm 0.8cm 0.7cm 0.9cm]{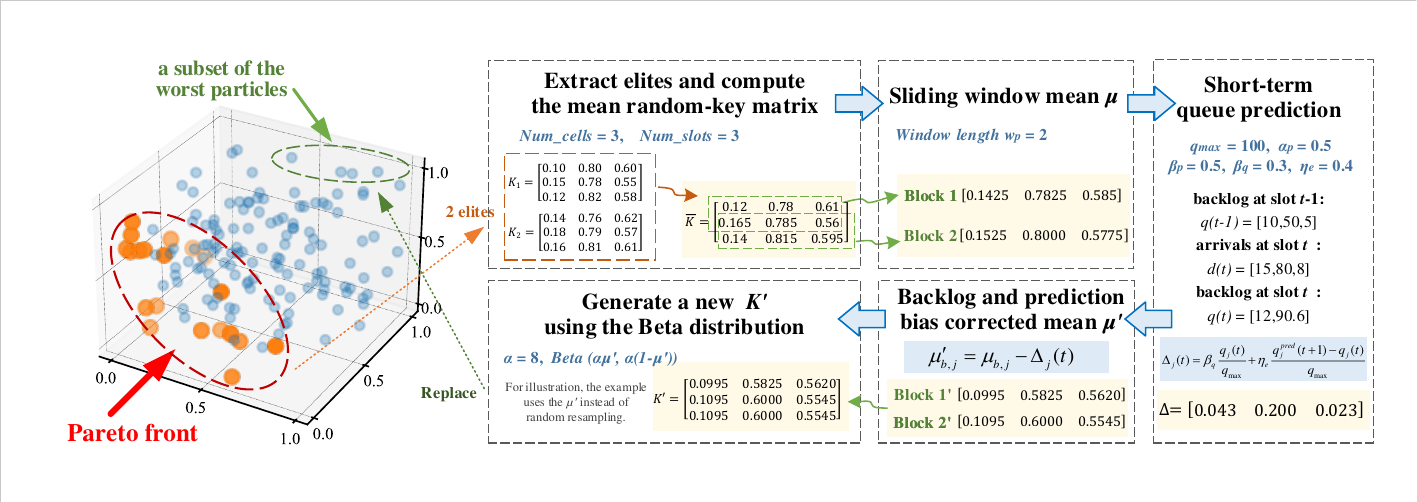} 
\caption{Illustrative example ($L=3, M=3, w_p=2$). The STP-PMD module first averages two elite key matrices and computes the sliding-window mean for each block--cell pair. It then evaluates the offset $\Delta_j(t)$ to obtain the corrected occupancy probability $\mu'_{b,j}$, based on which the Beta model is constructed and new keys $K'$ are sampled to replace the worst solution.}
\label{fig:PMD}
\vspace{-0.1cm}
\end{figure*}

To suppress clustering of population members in local optima and to exploit the spatiotemporal structure of the BHTP, we design a STP-PMD module, as shown in Fig.~\ref{fig:PMD}.
The module triggers every $T_{\mathrm{PMD}}$ generations.
First, select the top $\rho=20\%$ elite random-key matrices from the Pareto archive.
Then, apply exponential smoothing to each cell queue:
\begin{equation}\hat{q}_j(t)=\alpha_pd_j(t)+(1-\alpha_p)\hat{q}_j(t-1)\end{equation}
where $q_j(t)$ denote the backlog of cell $j$ at slot $t$;
$d_j(t)$ represents the actual arrival rate.
Then, generate a short-term queue prediction:
\begin{equation}\hat{q}_j^{\mathrm{pred}}(t+1)=\hat{q}_j(t)+\beta_p\left[\hat{q}_j(t)-\hat{q}_j(t-1)\right]\end{equation}
where $\alpha_p$ and $\beta_p$ are the coefficients of the module.

For the selected elite random-key matrices,
use an overlapping time window of length $w_p$ (in slots)
to compute the local mean $\mu$ of the key values.
Add a backlog- and prediction-based offset to obtain the corrected occupancy probability $\mu'_{b,j}$:
\begin{equation}
\mu'_{b,j}=\mu_{b,j}-\Delta_j(t)
\end{equation}
\begin{equation}
\Delta_j(t)=\beta_q\frac{q_j(t)}{q_{\max}}
+\eta_e\frac{\hat q^{\mathrm{pred}}_j(t+1)-q_j(t)}{q_{\max}},\ \forall j,
\end{equation}
where $\beta_q$ and $\eta_e$ are weighting coefficients, 
$q_{\max}$ is the backlog normalization constant,
and $\Delta_j(t)$ is the backlog-and-prediction offset for cell $j$.
This offset biases the sliding-window mean toward cells with larger backlog and higher predicted load.
We then build a probabilistic model $X\sim \mathrm{Beta}(\alpha \mu',\, \alpha(1-\mu'))$ to guide resampling, where $\mathrm{Beta}(a,b)$ denotes the Beta distribution on $(0,1)$ with $a>0$ and $b>0$.
Its probability density function (PDF) is
\begin{equation}
f_X(x;a,b)=\frac{1}{\mathrm{B}(a,b)}x^{a-1}(1-x)^{b-1},\quad 0<x<1.
\end{equation}
This parameterization satisfies $\mathbb{E}[X]=\mu'$ and $\mathrm{Var}(X)=\mu'(1-\mu')/(\alpha+1)$; thus the model carries spatiotemporal priors through $\mu'$, while $\alpha$ controls the concentration around $\mu'$.

Finally, we sample a fraction $\xi=20\%$ of new random keys from the model
to refresh inferior particles, and apply small perturbations.
These samples replace particles with the worst fitness and the highest crowding. During decoding, the algorithm prefers cells that have delivered gains in the corresponding slots historically. It also preserves temporal smoothness and population diversity. As a result, it is easier to produce BHTP that satisfy hotspot preference and fairness constraints.
The short-term prediction only biases the probabilistic resampling, rather than directly determining the final schedule. Therefore, prediction errors mainly affect candidate-generation efficiency, while the final solutions are still screened through multi-objective evaluation and archive update. Moreover, under rolling online replanning, newly observed traffic is incorporated in subsequent updates, which helps reduce the impact of abrupt traffic changes.

\begin{algorithm}[t]
  \caption{Short-term Prediction Probability Model Driven Resampling (STP-PMD)}
  \label{alg:STP-PMD}
  \KwIn{$P=\{K_1,\dots,K_{|P|}\}$; archive $A$; queues $q_j(t)$; 
  $w_p$;
  number of slots $L$;
  number of cells $M$;
  number of blocks $N_{\mathrm{blk}}$;
  elite fraction $\rho$; 
  refresh fraction $\xi$; $\alpha_p,\beta_p,\beta_q,\eta_e,\alpha,\sigma_i$; $q_{\max}$}
  \KwOut{Updated $P$ and $A$}
  \BlankLine
  $g\gets 0$; $N_{\mathrm{blk}}\gets L-w_p+1$\;
  \While{termination criterion not met}{
    \textsc{EvaluateFitness}$(P)$\; \textsc{UpdatePbest}$(P)$\; \textsc{UpdateArchive}$(A)$\;
    \If{$g$ $\bmod$  $T_{\mathrm{PMD}} = 0$}{
      $E$$\gets$ top $\rho|P|$ elites from $A$ \;
      \ForEach{cell $j$}{
        $\hat q_j(t)\gets \alpha_p d_j(t)+(1-\alpha_p)\hat q_j(t-1)$; $\hat q^{\mathrm{pred}}_j(t+1)\gets \hat q_j(t)+\beta_p\,[\hat q_j(t)-\hat q_j(t-1)]$\;
      }
      \For{$b\gets 1$ \KwTo $N_{\mathrm{blk}}$, $j=1\dots M$}{
  $\mu_{b,j}\gets \mathrm{mean}\big(E\ \text{rows }[b{:}b{+}w_p{-}1],\ \text{col }j\big)$\;
  $\Delta_j(t)\gets \beta_q\,\dfrac{q_j(t)}{q_{\max}}+\eta_e\,\dfrac{\hat q^{\mathrm{pred}}_j(t+1)-q_j(t)}{q_{\max}}$\;
  $\mu_{b,j}'\gets \mu_{b,j}-\Delta_j(t)$\;
}
      \For{$b\gets 1$ \KwTo $N_{\mathrm{blk}}$, $j=1\dots M$}{$a_{b,j}\gets \alpha \mu_{b,j}'$; $b_{b,j}\gets \alpha(1-\mu_{b,j}')$;}
      $S_{\mathrm{new}}\gets \emptyset$\;
      \For{$k\gets 1$ \KwTo $\lceil \xi|P|\rceil$}{
        $K'\gets \textsc{SampleBeta}$ $(a_{b,j},b_{b,j})$ + $\mathcal N(0,\sigma_i^2)$ \;
        $S_{\mathrm{new}}\gets S_{\mathrm{new}}\cup\{K'\}$\;
      }
      $\textit{Worst} \gets \textsc{SelectWorst}\big(P,\lceil \xi|P|\rceil\big)$\; \textsc{Replace}$(\textit{Worst},S_{\mathrm{new}})$\;
    }
    $g\gets g+1$\;
  }
  \Return{$P, A$}\;
\end{algorithm}

The pseudocode of the STP-PMD module is described in Algorithm~\ref{alg:STP-PMD}.
Lines 1–5 perform initialization and the standard iterative loop. When the trigger condition is met, the PMD branch is executed.
Lines 6–7 select a fraction $\rho$ of elites by nondominated rank and crowding.
Lines 8–9 compute the short-term queue predictions.
Lines 10–12 compute window means and add offsets. It then maps these occupancies to Beta distribution parameters.
Given this prior, Lines 13–18 sample new random keys and replace inferior solutions.

The time complexity of STP-PMD is summarized as follows. A single trigger incurs four main costs: elite selection $O(|A|\log|A|)$;
window statistics $O(N_{\mathrm{blk}}Mw_{p})$;
sampling and assembling $\xi|P|$ key matrices $O(\xi|P|LM)$;
selection and replacement of inferior solutions $O(|P|\log|P|)$.
Over $G$ generations, with a trigger every 
$T_{\mathrm{PMD}}$ generations, the total complexity is:
\begin{equation}O{\left(\frac{G}{T_{\mathrm{PMD}}}\left[\xi|P|\cdot LM+(|A|+|P|)\log|P|+N_{\mathrm{blk}}Mw_p\right]\right)}\end{equation}

\textit{Example Analysis:} As shown in Fig.~\ref{fig:PMD}, we set $L=3$ time slots, $M=3$ cells, and window length $w_p=2$. First, the algorithm takes two elite random-key matrices and computes their element-wise mean.
It then applies a sliding-window mean over $block \times cell$; for Block 1, it uses rows 1–2.
Given the backlogs and arrivals, it computes the correction $\Delta$ via the prediction formula and updates the block values. 
With $\alpha=8$, it builds a Beta model.
For reproducibility, it uses the Beta expectation instead of random sampling and generates a new random-key matrix $K'$.
Finally, the worst solution is replaced by the newly generated matrix $K'$.

\begin{figure}[t]
\centering  
\includegraphics[width=0.9\linewidth, clip, 
trim=1cm 1cm 1cm 1cm]{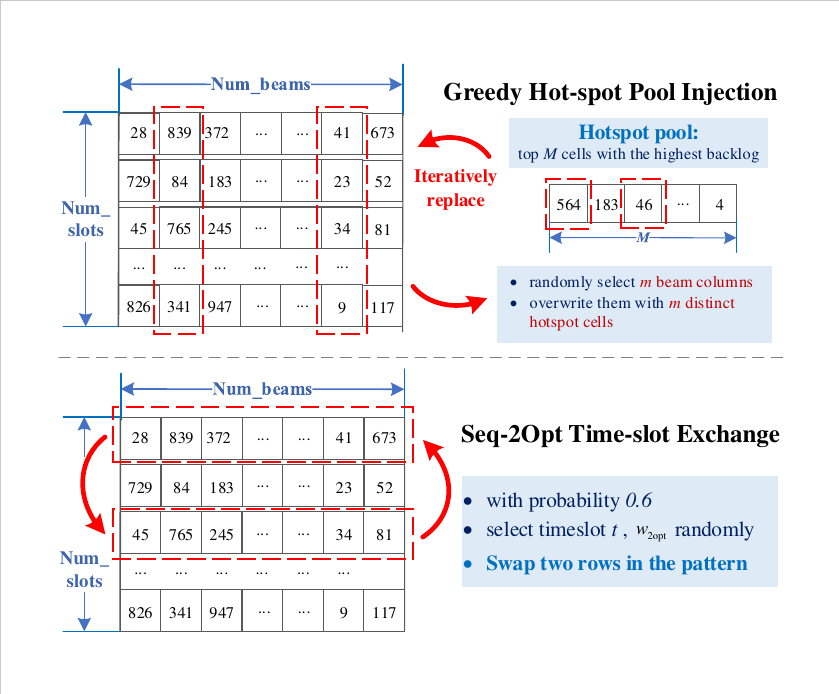} 
\caption{LR module. 
Greedy injection repairs the worst $\eta_{\rm GI}=25\%$ of particles using cells from the top-$M$ backlog. Seq-2Opt swaps two slot patterns within a span window with probability $p_{\rm 2opt}=0.6$, and keeps the swap only if throughput increases.
}
\label{fig:LRmodule}
\vspace{-0.2cm}
\end{figure}

\subsection{Local Refinement Module (LR)}

To avoid premature convergence caused by population members trapped in suboptimal regions, we introduce the LR module, as shown in Fig.~\ref{fig:LRmodule}.
It comprises two operators:
greedy injection and Seq-2Opt timeslot exchange.
The LR module activates every $T_{\mathrm{LR}}$ generations.
In the greedy injection step, particles are ranked by instantaneous
throughput, and the worst $\eta_{\rm GI}=25\%$ are selected for repair.
Then, build a hotspot pool from the top 
$M$ high-load cells. 
Finally, randomly choose $m$ beam columns and overwrite each full column in the BHTP with mutually distinct hotspot cells from the pool.

Next, the Seq-2Opt operator is executed with probability
$p_{\rm 2opt}=0.6$ to locally adjust the BHTP slot order.
Within a window of length $w_{\mathrm{2opt}}$ (in slots), randomly swap two BHTP rows.
$w_{\mathrm{2opt}}$ is the row-swap neighborhood size, and each row represents one slot's illumination pattern. 
The swap is accepted only if the resulting throughput increases.

Set the fitness of all modified particles to \(+\infty\) to force reevaluation in the next generation and keep them in the evolutionary process.
This rapidly clears hotspots and advances the Pareto front without sacrificing exploration.

\subsection{Multi-Scale Perturbation Module (MSP)}

The MSP module applies coarse- and fine-scale perturbations at the population level, together with a $\sigma_i$-adaptive perturbation at the individual level, where $\sigma_i$ denotes the private perturbation step size of particle $i$.
It operates directly on the BHTP random-key matrix at the end of each generation.
At initialization, 20\% of the particles receive a coarse step size, while the remaining particles use a fine step size. Then, Gaussian perturbation is applied in each generation:
\begin{equation}
\widetilde{key}_{t,c} = (key_{t,c} + \mathcal{N}(0,\sigma_i)) \bmod 1
\end{equation}
With probability 0.5, the perturbed particle is averaged with an individualized guiding solution to avoid excessive drift.
To adapt the perturbation scale, the 1/5th success rule is used within a window of length $w_{\sigma}$.
A success is recorded when particle $i$ updates its personal best within the current window, and the resulting success rate is used to adjust $\sigma_i$.
Thus, MSP enables an adaptive transition from coarse exploration to fine-grained refinement.

\subsection{Convergence and Computational Complexity}
Since the studied BHTP is a high-dimensional discrete multi-objective optimization problem, we design Aidos as a population-based metaheuristic
rather than a deterministic solver.
We discuss its convergence behavior from three aspects:
elitist archive preservation, non-worsening local refinement, and bounded guided perturbation, as summarized in the following propositions.

\textbf{Proposition 1: Elitist archive preservation.}
Let $A_t$ denote the external archive at generation $t$. 
If $A_{t+1}$ keeps the non-dominated solutions from the current archive and the newly generated candidates, with diversity-based truncation when needed, then previously found high-quality solutions cannot be replaced by dominated ones.

This proposition shows that the archive update is elitist. Therefore, high-quality non-dominated solutions can be preserved throughout subsequent iterations.

\textbf{Proposition 2: Non-worsening local refinement.}
If a local refinement operator accepts a new solution only when the acceptance criterion improves, then each accepted refinement step is non-worsening under that criterion.

In Aidos, Seq-2Opt in the LR module is accepted only when the throughput of the swapped pattern exceeds that of the original pattern. Thus, local refinement is a directed improvement rather than a random perturbation.

\textbf{Proposition 3: Bounded perturbation scale.}
If the perturbation step size of each particle is kept within a predefined bounded range, then the update magnitude of the MSP module remains bounded in every generation and cannot diverge due to unbounded step growth.

Since the step size is always bounded, the search gradually shifts from exploration to refinement.

These properties do not establish convergence to the global Pareto-optimal set. They suggest that Aidos is a structured iterative process.
This is also consistent with the empirical convergence in Section V.

The per-generation cost of Aidos consists of four parts:
TARK decoding and fitness evaluation for all $|P|$ particles, the MSP update, and the periodically triggered STP-PMD and LR modules.
The complexity of TARK decoding has been given above, and the total cost of fitness evaluation is denoted by $C_{\mathrm{eval}}$.
MSP updates all random-key matrices, with complexity $O(|P|\cdot LM)$.
The complexity of STP-PMD is given by Equation (17), and the amortized complexity of LR is $O(C_{\mathrm{LR}}/T_{\mathrm{LR}})$.
Therefore, the overall complexity is:
\begin{equation}
\label{eq:total_per_gen_complexity}
\begin{aligned}
O\Bigg(&|P| \cdot LM\log M + C_{\mathrm{eval}} + |P| \cdot LM
+ \frac{C_{\mathrm{LR}}}{T_{\mathrm{LR}}} \\
&+ \frac{\xi |P| \cdot LM + (|A|+|P|)\log |P| + N_{\mathrm{blk}}Mw_p}
{T_{\mathrm{PMD}}}
\Bigg)
\end{aligned}
\end{equation}

\section{EVALUATION}

\subsection{Experiment Setup}

\textit{Simulation Scenario:} 
This study evaluates Aidos on the second-generation Starlink constellation to verify its effectiveness in large-scale satellite networks.
The constellation uses circular orbits at 550 km. It comprises 72 orbital planes with 22 satellites per plane, for a total of 1,584 satellites.
Each satellite carries 48 independent downlink spot beams.
Each beam provides a peak capacity of 200 \,Mbps, as illustrated in TABLE~\ref{tab:parameters}. 

\begin{table}[!h]
\renewcommand{\arraystretch}{1.2}
\caption{Simulation Parameters}
\label{tab:parameters}
\centering
\begin{tabular}{ll}
\toprule
\textbf{PARAMETERS}   &  \textbf{VALUES} \\
\midrule
\textbf{COMMUNICATION PARAMETERS} &  \\
Satellite altitude $H$  &  550 km \\
Number of orbital planes  & 72 \\
Number of satellites per orbital plane  & 22\\
Orbit inclination  &  53°\\
Number of downlink beams per satellite & 48\\
Total capacity per satellite & 20 Gbps\\
Per-beam capacity & 200 Mbps\\
Carrier frequency $f_c$  & 20 GHz\\
3-dB beamwidth $\theta_{3\mathrm{dB}}$ & 0.15°\\
Satellite transmit $\mathrm{EIRP}$  & 60 dBW\\
H3 res  & 5\\
Time slot duration $T_{\text{slot}}$  & 30 ms\\
Total duration of BHTP & 600 ms\\

\midrule
\textbf{OUTER-LOOP SWARM PARAMETERS} &  \\
Number of particles $N_p$ & 20 \\
Number of iterations $N_{\mathrm{iter}}$ & 150 \\
Cognitive acceleration coefficient $c_1$ & 1.5 \\
Social acceleration coefficient $c_2$ & 1.5 \\
Inertia weight $w_{\mathrm{in}}$ & 0.7 \\
\midrule
\textbf{MODULE PARAMETERS} &  \\
Elite fraction $\rho$ & 0.2 \\
Refresh fraction $\xi$ & 0.2 \\
Beta concentration $\alpha$ & 4 \\
Window length $w_p,w_{\sigma}$ & 4,30 \\
Backlog weight $\beta_q$ & 0.6 \\
Prediction weight $\eta_e$ & 0.4 \\
Greedy injection fraction $\eta_{\rm GI}$ & 0.25\\
Seq-2Opt probability $p_{\rm 2opt}$ & 0.6 \\

\bottomrule
\end{tabular}
\end{table}

\begin{figure}[h]
\centering
\includegraphics[width=0.9\linewidth]{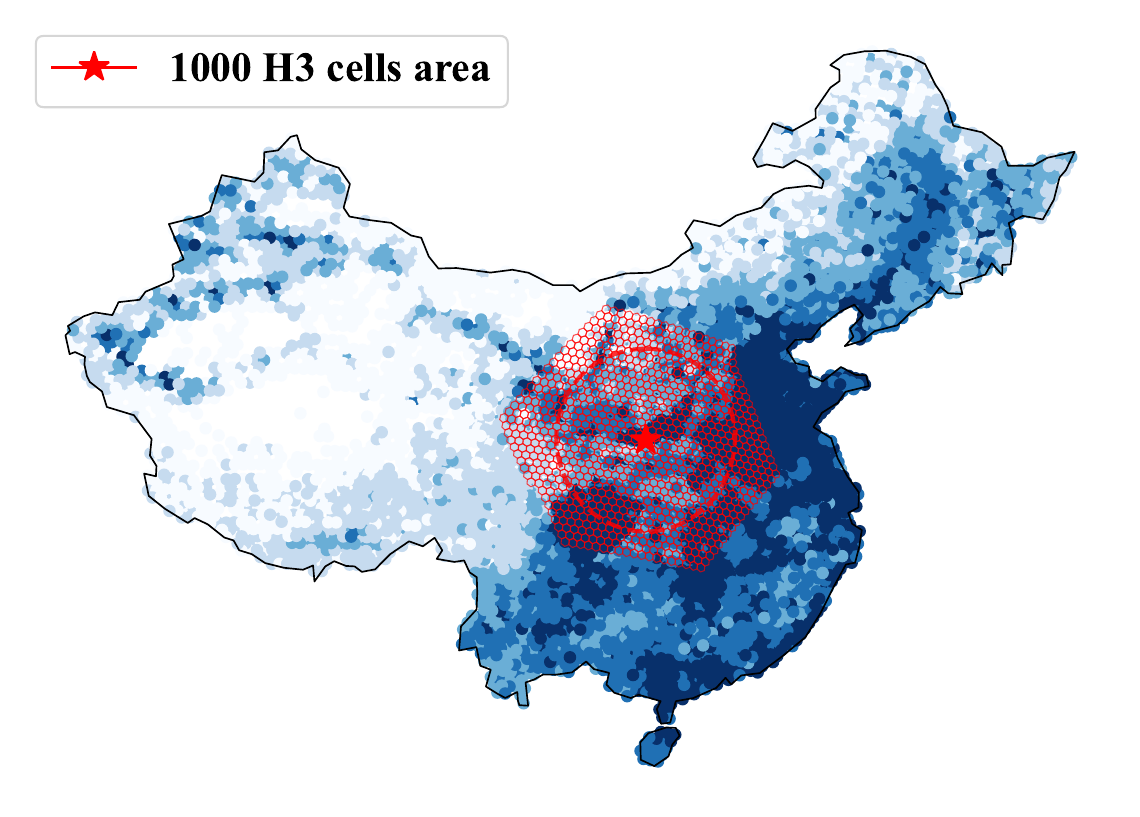} 
\caption{Kontur population map (H3 Res 5, China); red: 1000-cell coverage area.}
\label{fig:pop}
\end{figure}

The preset ratios in Table~II are module-level hyperparameters of Aidos.
The elite fraction $\rho=0.2$ keeps STP-PMD focused on high-quality Pareto
solutions while retaining enough samples for stable slot-cell selection
statistics. The refresh fraction $\xi=0.2$ injects Beta-resampled
candidates biased toward high-demand cells. In LR, greedy injection
repairs the worst $\eta_{\rm GI}=0.25$ particles to improve
low-throughput BHTP, and Seq-2Opt uses $p_{\rm 2opt}=0.6$ for local
BHTP-row swaps that refine the slot order.

\begin{figure}[!t]
\centering
\includegraphics[width=0.9\linewidth]{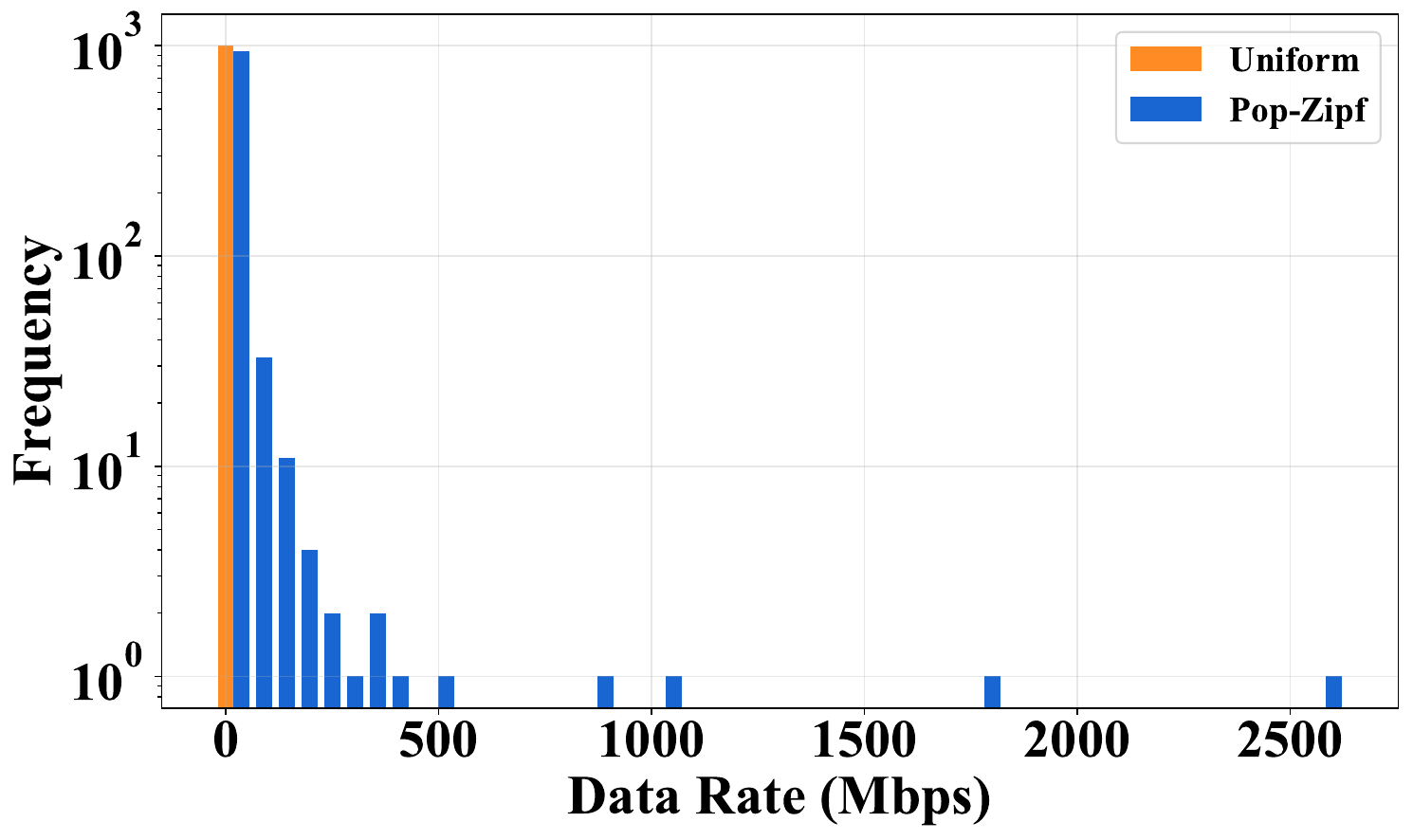} 
\caption{Traffic Distribution Histogram. 
Under uniform traffic, most cell demands do not exceed 10 Mbit/s, whereas the population–Zipf model exhibits a pronounced long-tail distribution, with a minority of hotspot cells exceeding several hundred Mbit/s.}
\label{fig:traffic}
\vspace{-0.3cm}
\end{figure}

\textit{Dataset:}
Due to the absence of a public satellite-network traffic dataset, we generate a dataset based on the Kontur world population grid and the traffic-modeling principles in 3GPP TR 38.821.
Ground cells within satellite coverage are derived using the H3 hexagonal grid library.
Each cell’s initial traffic demand combines Kontur population data with Zipf-law weighting. We then add independent Poisson increments at each time slot to emulate bursty arrivals.
Fig.~\ref{fig:pop} shows a population heatmap of mainland China on an H3-5 grid. Fig.~\ref{fig:traffic} shows the traffic distribution histogram. 
The uniform baseline concentrates near 0–10 Mbps. The population–Zipf case exhibits a clear long tail, with a small number of cells reaching several hundred Mbps.
These figures confirm strong load heterogeneity: hotspot cells consume a large share of capacity. The dataset approximates real-world conditions and provides a credible basis for evaluating BH schedulers.

\textit{Overall Performance:}
We evaluate throughput and average real-time packet delay over the entire BH cycle.
To quantify Aidos under different traffic intensities, we set the total satellite capacity to $D$ and evaluate performance with ground demand $[0.2, 0.4, 0.6, 0.8, 1.0]\times D$. We report the trends of the performance metrics.

The proposed Aidos is benchmarked against five baseline BH schemes:

\begin{enumerate}[label={(\arabic*)}]
  \item \textbf{Random-BH}: Randomly selects 
$K_\mathrm{max}$ cells each time slot, incurring negligible cost and serving as a lower-bound benchmark.
  
  \item \textbf{Greedy-BH}: Select the top $K_\mathrm{max}$ cells with the highest backlog each time slot to maximize instantaneous throughput.
  
  \item \textbf{MOGA-BH}: The BH schedule is encoded as chromosomes. Genetic operations including selection, crossover, and mutation are applied to evolve solutions across generations.
  
  \item \textbf{SA-BH}: Simulated annealing perturbs the current solution and probabilistically accepts worse moves under a temperature schedule to evade local optima; precise cooling-schedule tuning remains essential.

  \item \textbf{MADRL-BH}: According to \cite{lin2024satellite}     
  , the MADRL algorithm is used to compute the illumination pattern in each time slot.
\end{enumerate}

\begin{figure}[!t]
\centering  
\includegraphics[width=0.9\linewidth, clip, 
trim=0.1cm 0.1cm 0.1cm 0.1cm]{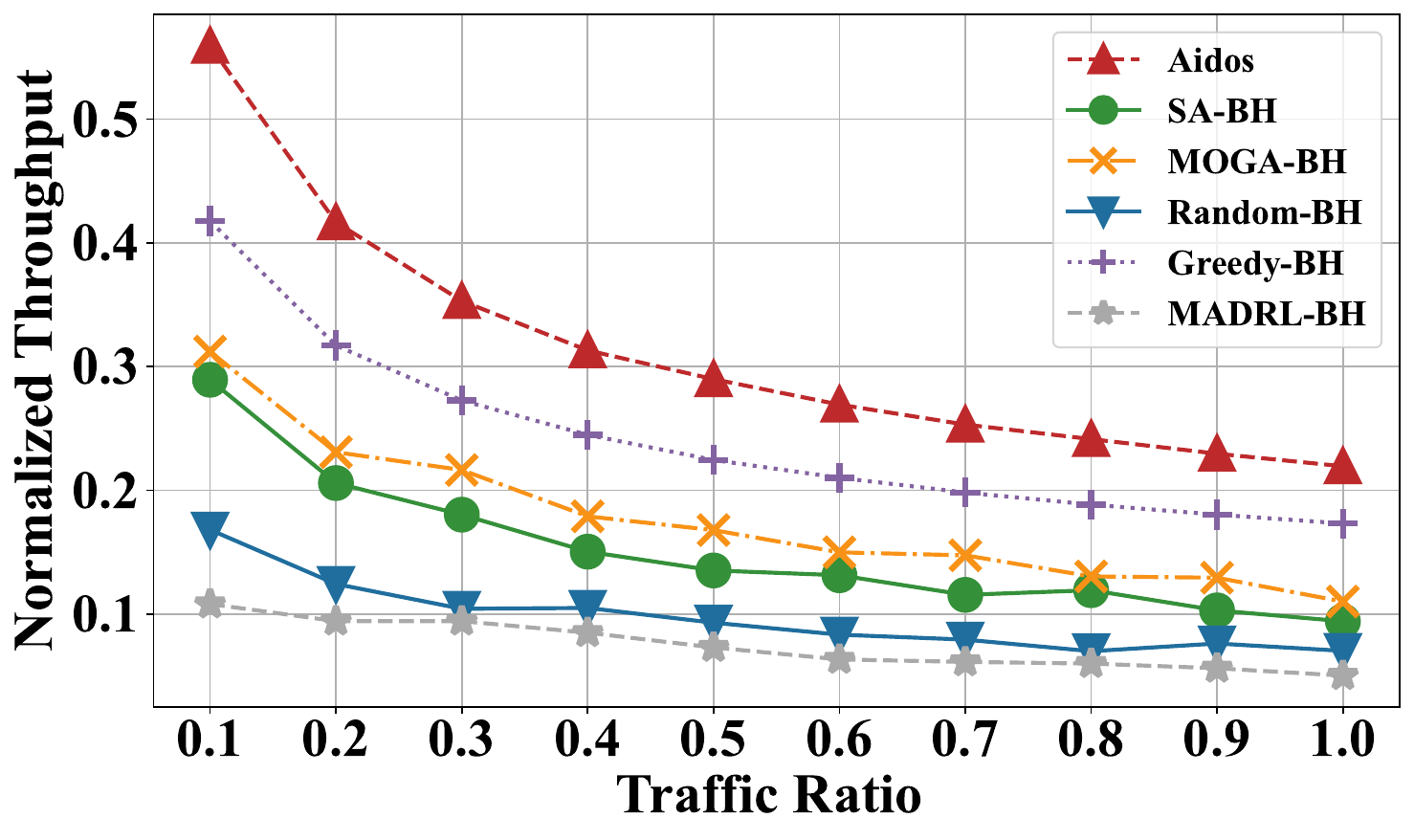} 
\caption{Throughput performance trends of multiple algorithms under varying ground-traffic loads. Among the compared algorithms, Aidos achieves the highest normalized throughput.}
\label{fig:100throughput}
\end{figure}

\begin{figure}[!t]
\centering  
\includegraphics[width=0.9\linewidth, clip, 
trim=0.1cm 0.1cm 0.1cm 0.1cm]
{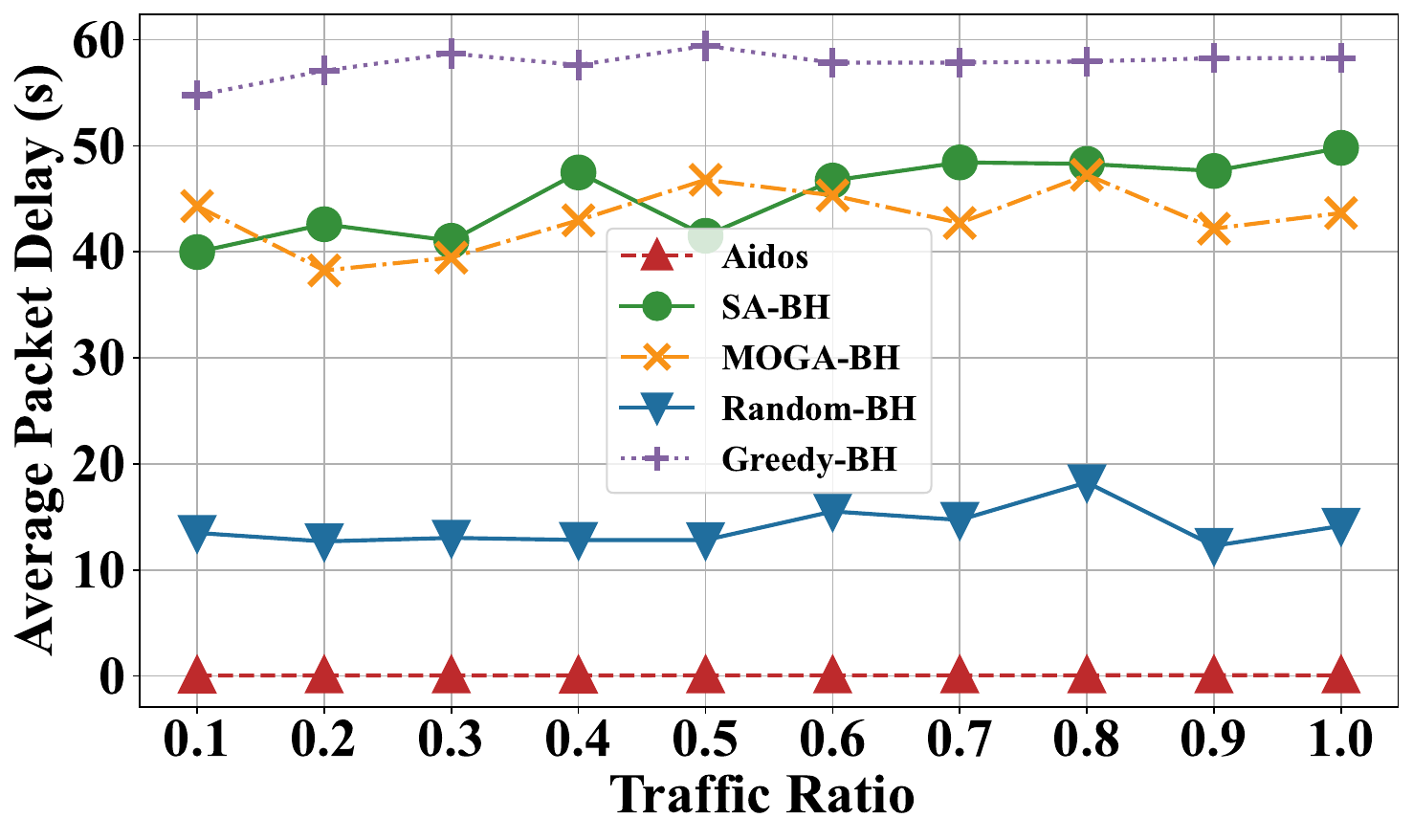} 
\caption{Delay performance trends of multiple algorithms under varying ground-traffic loads. Among the compared algorithms, Aidos achieves the lowest average packet delay.}
\label{fig:100delay}
\vspace{-0.2cm}
\end{figure}

\begin{figure*}[!t]
    \centering
    \subfloat[TARK module\label{fig:TARK}]{%
        \includegraphics[width=0.32\textwidth, 
        clip, trim=0.2cm 0.1cm 0.2cm 0.1cm]{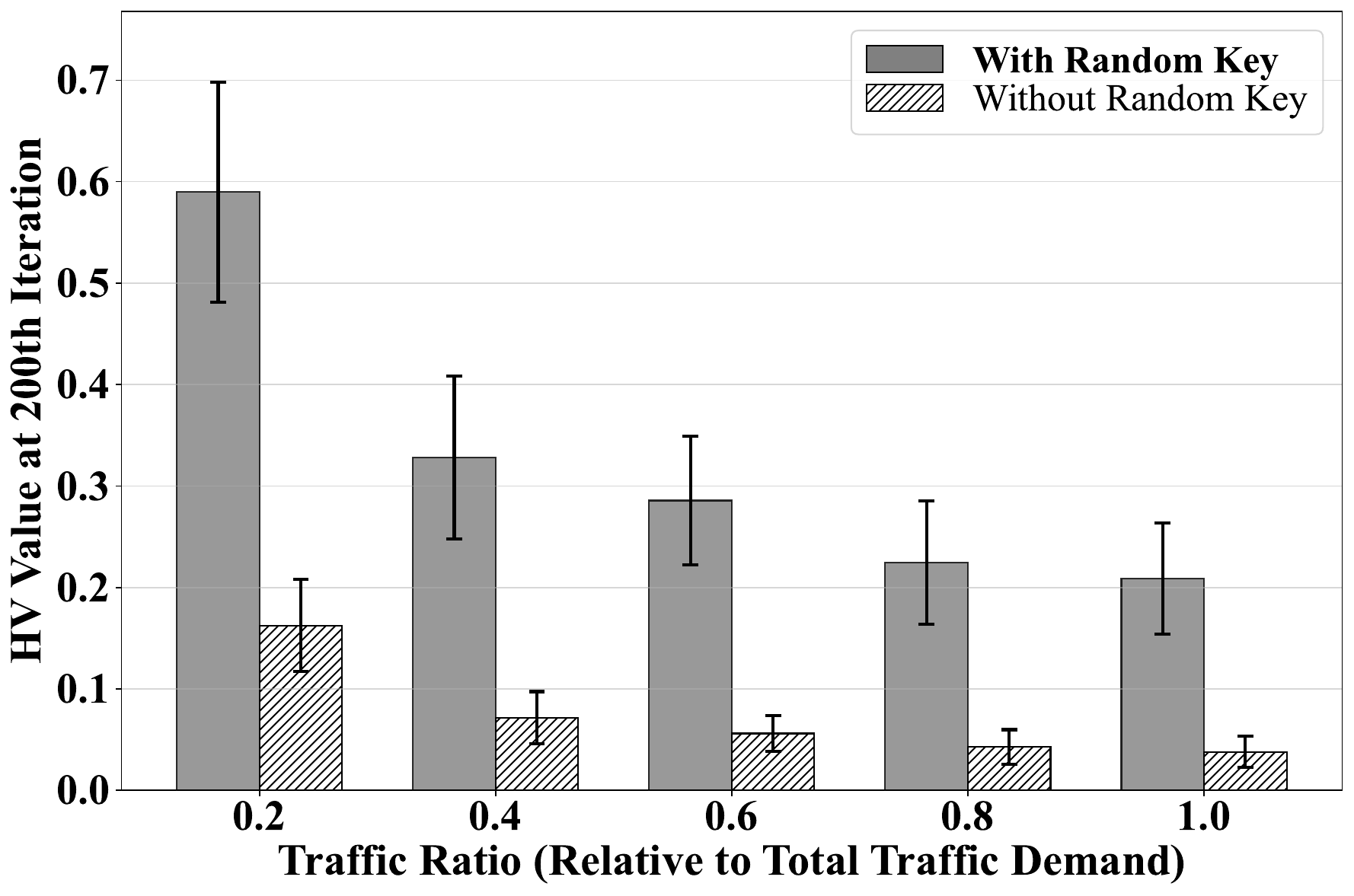}
    } \hfill
    \subfloat[STP-PMD module\label{fig:STP-PMD}]{%
        \includegraphics[width=0.32\textwidth, 
        clip, trim=0.2cm 0.1cm 0.2cm 0.1cm]{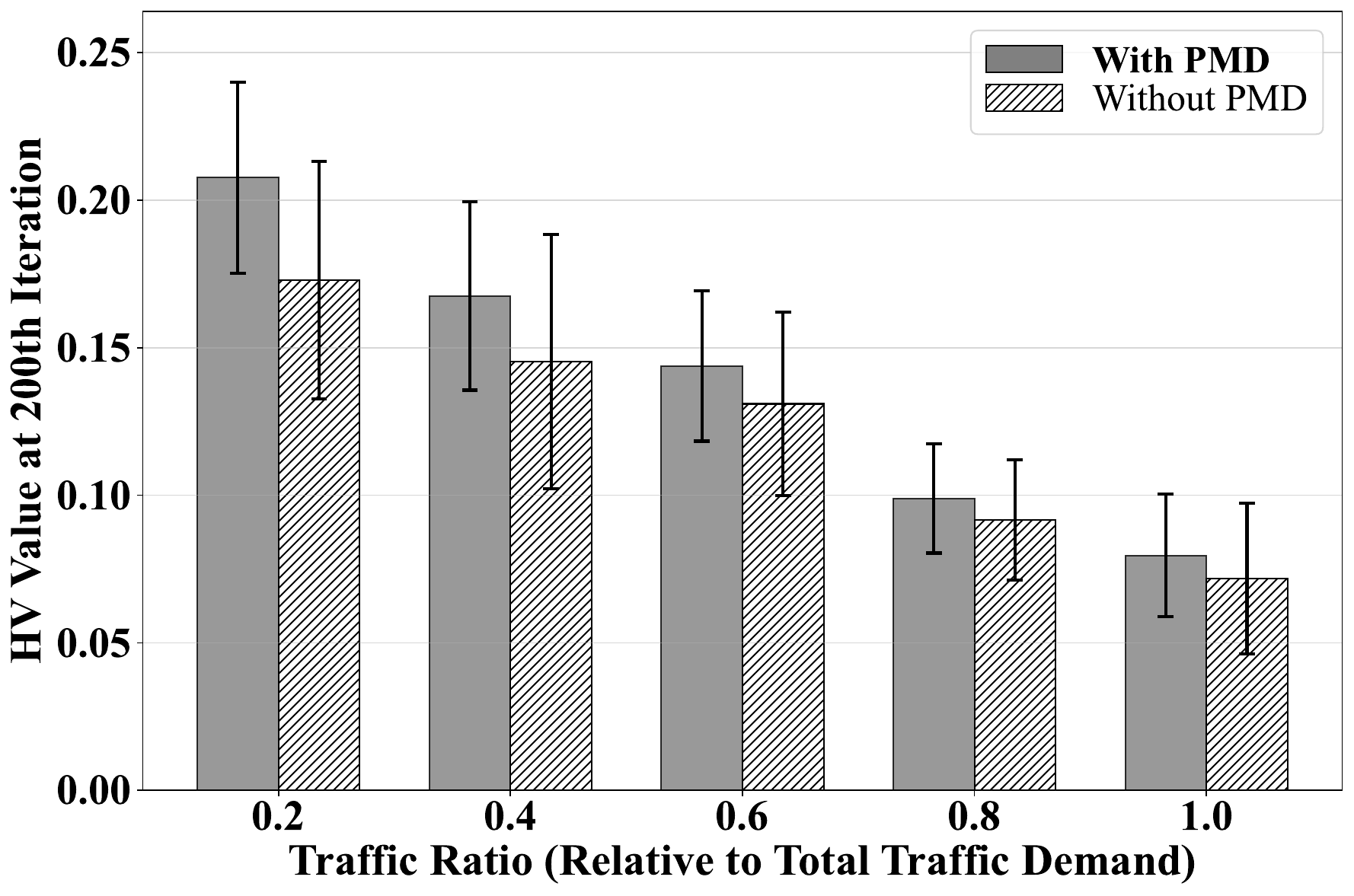}
    } \hfill
    \subfloat[MSP module\label{fig:MSP}]{%
        \includegraphics[width=0.32\textwidth, 
        clip, trim=0.2cm 0.1cm 0.2cm 0.1cm]{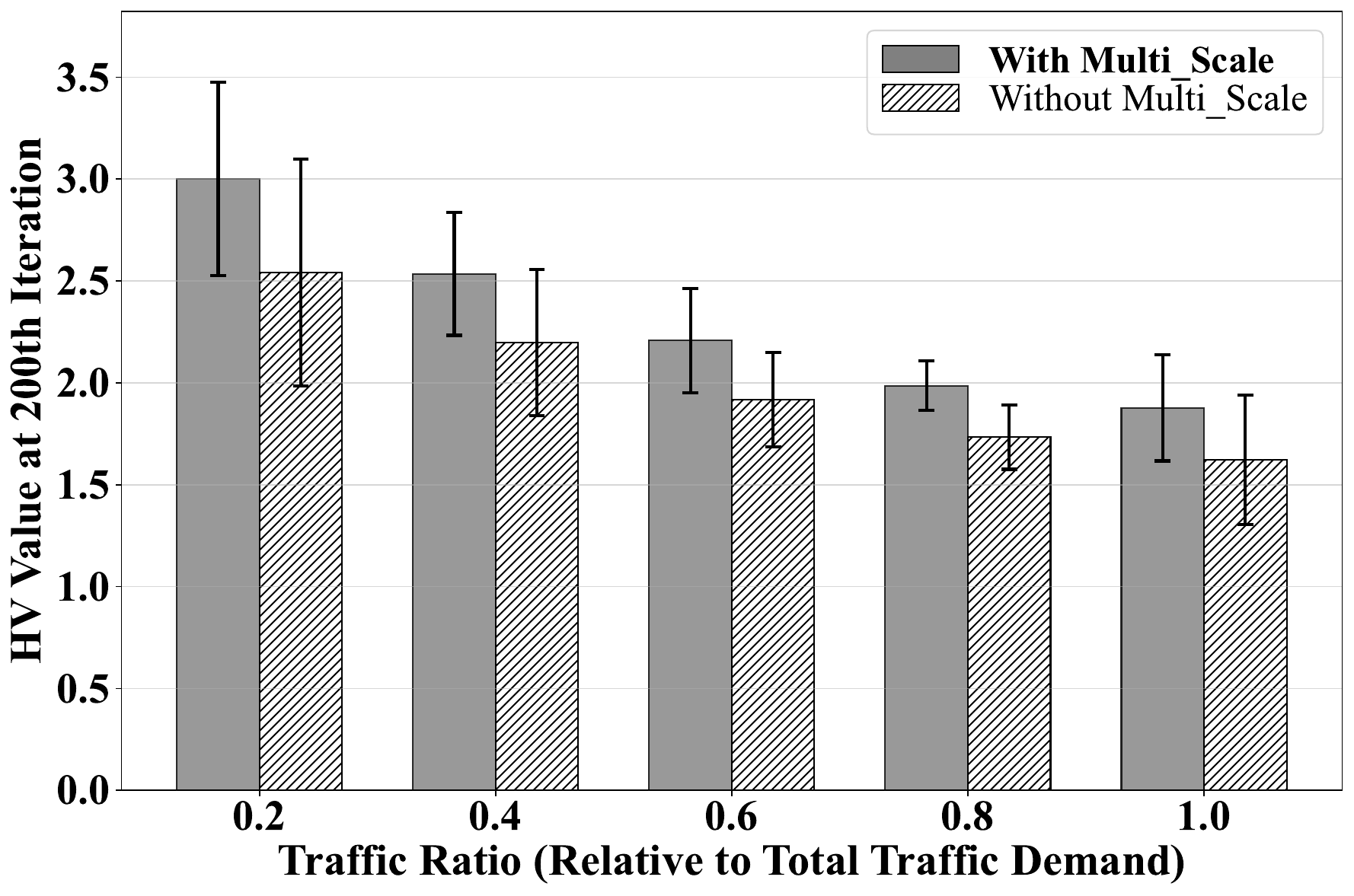}
    }
    \caption{ The ablation study results of TARK, STP-PMD and MSP modules. Experimental results indicate that all three modules effectively broaden the optimization algorithm’s solution space, thereby enhancing the overall HV performance. }
    \label{fig:ablation}
\end{figure*}

\begin{figure}[t]
    \centering
    \subfloat[Delay\label{fig:LRdelay}]{
        \includegraphics[width=0.24\textwidth]{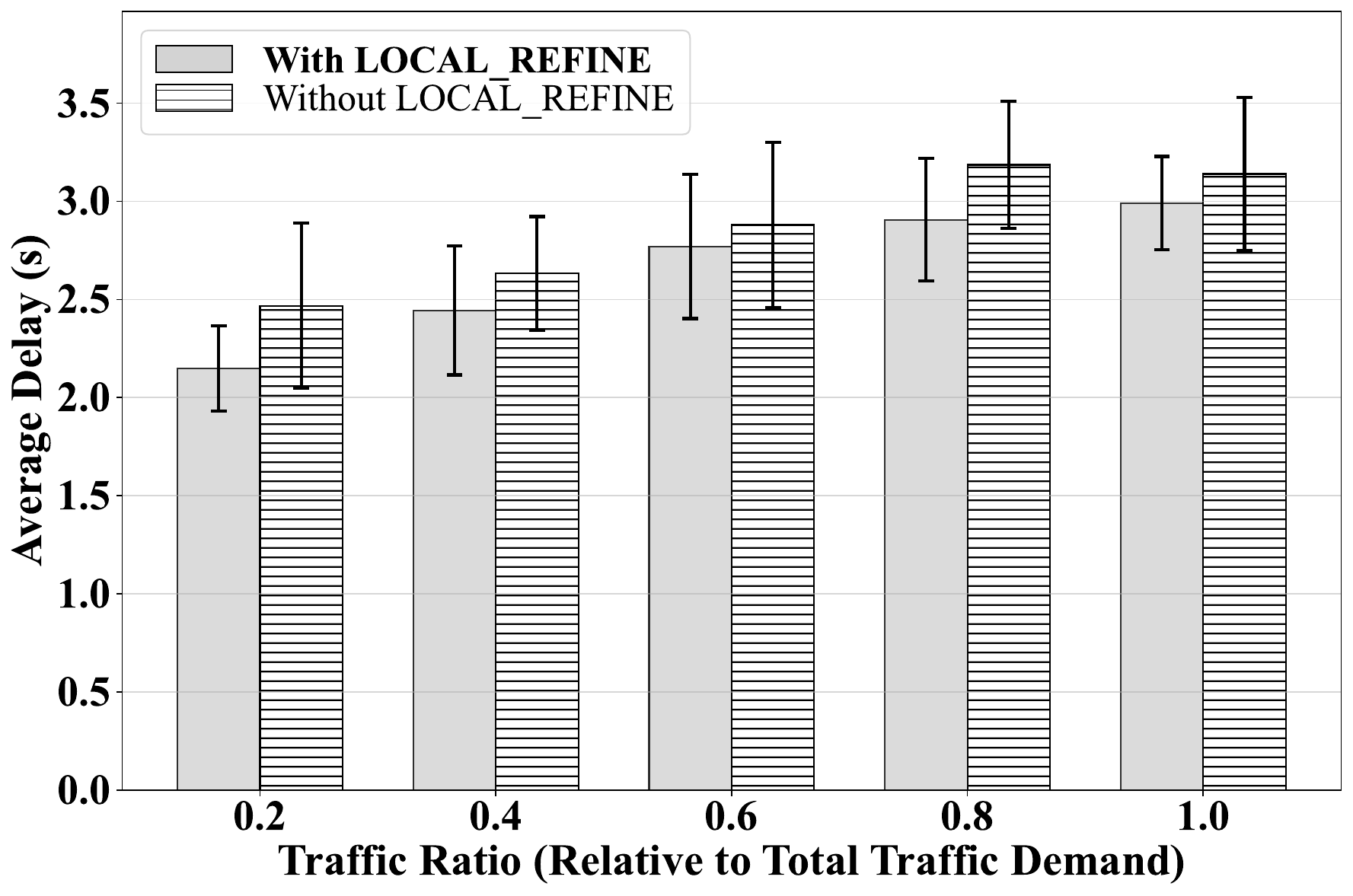}
    }
    \subfloat[HV\label{fig:LRHV}]{
        \includegraphics[width=0.24\textwidth]{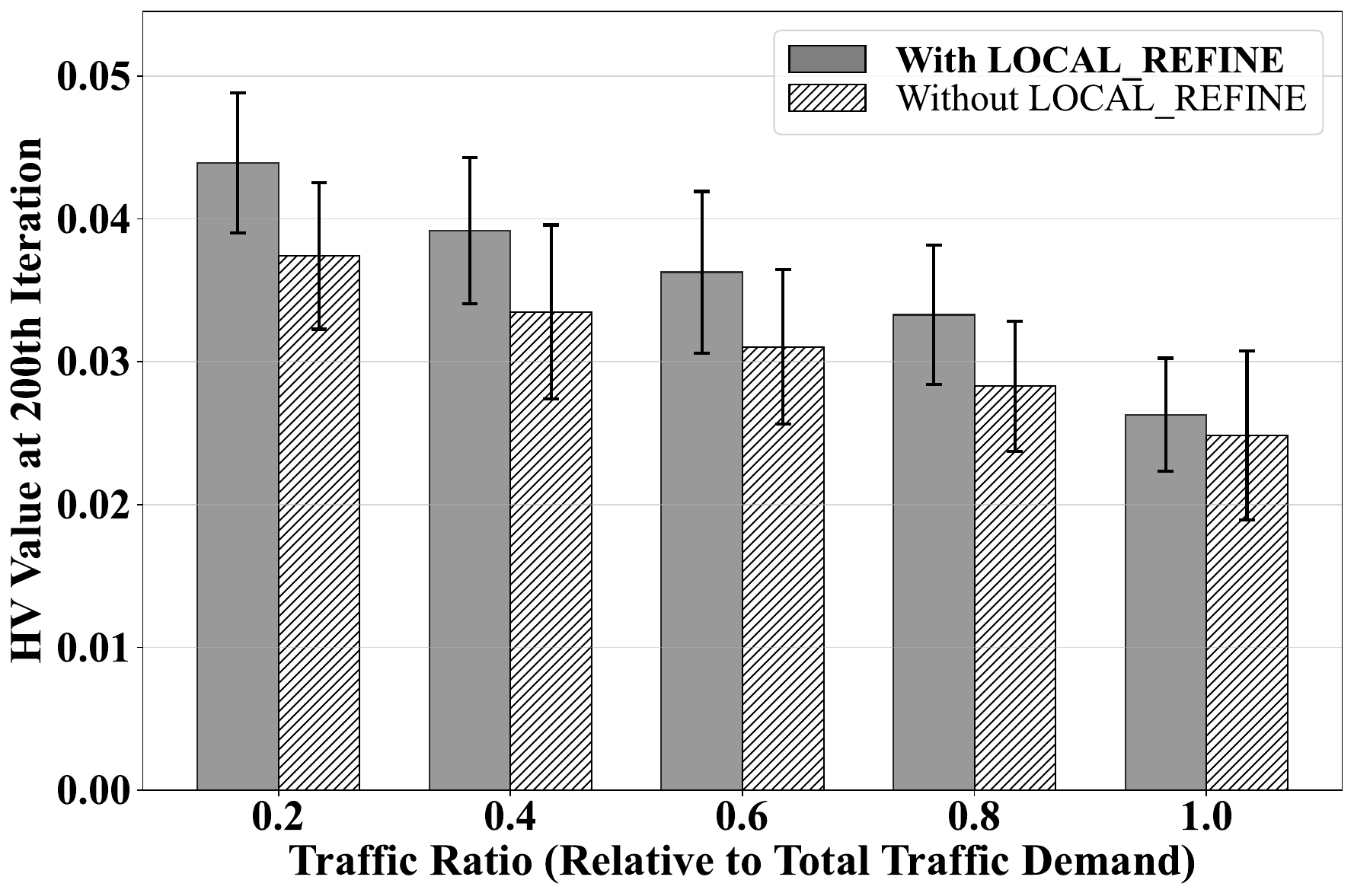}
    }
    \caption{The ablation study results of LR module. With the LR module enabled, latency is reduced by approximately 7\%, while the HV metric improves by 5\%-10\%. }
    \label{fig:LR}
\end{figure}

\textit{Ablation Study:} 
To quantify the independent contribution of four key mechanisms in Aidos, we conduct ablation studies on the following modules: TARK, STP-PMD, LR, and MSP. 
We disable one module at a time while keeping the remaining pipeline and hyper-parameters unchanged. This procedure yields four controlled variants, and the full Aidos serves as the baseline. All variants run on the same traffic scenario with ten random seeds. We record the main performance metrics and the hyper-volume (HV) curves.
  
\textit{Convergence:}
Metaheuristic algorithms improve solution quality through iterative optimization.
In practice, convergence is constrained by the available computation time.
Consequently, we evaluate SA-BH, MOGA-BH, and Aidos over 300 iterations.
We report the mean and the 95\% confidence interval across 10 independent runs. All experiments use the same traffic load of 10 Gbps.

\begin{figure}[t]
\centering  
\includegraphics[width=0.9\linewidth, clip, 
trim=0.1cm 0.1cm 0.1cm 0.1cm]
{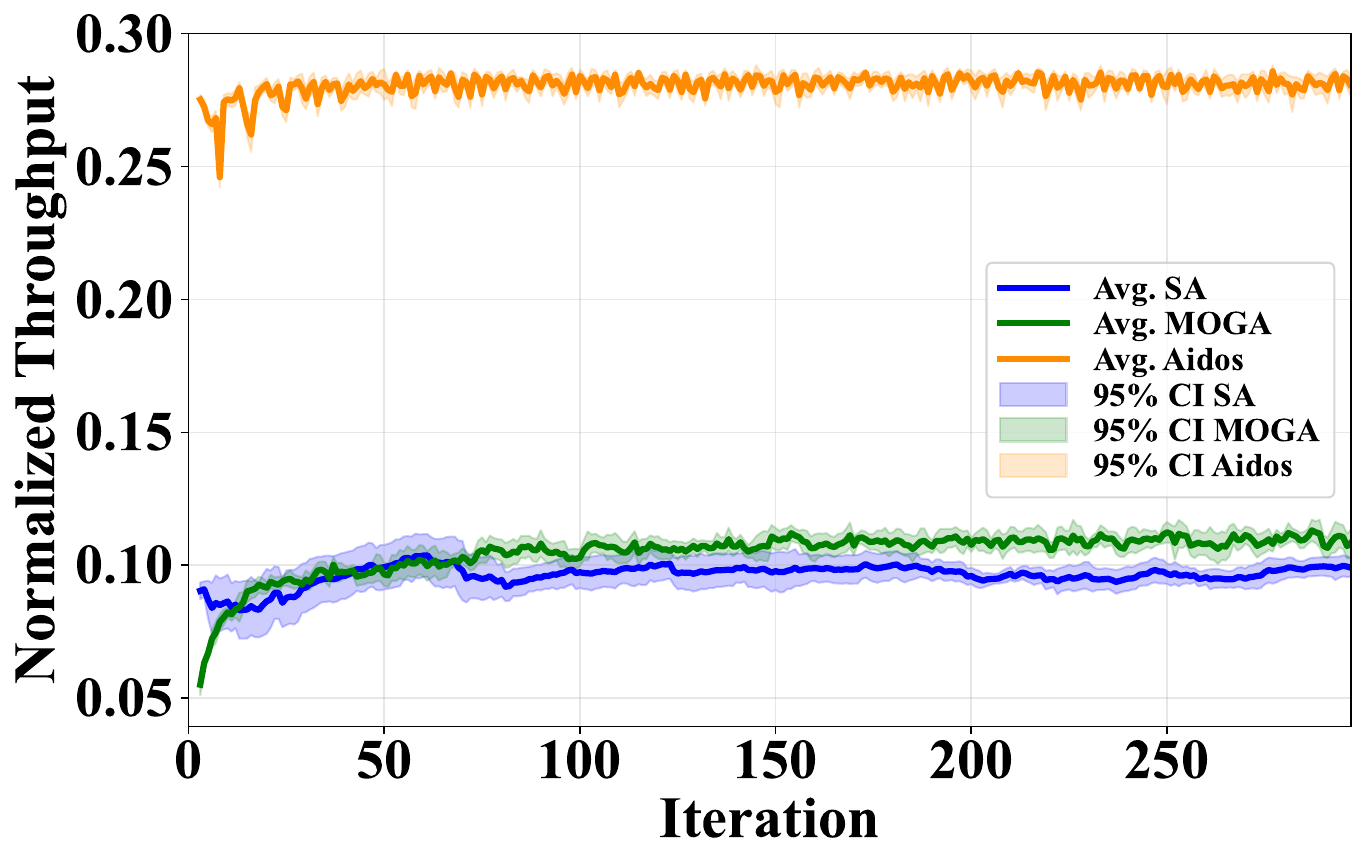} 
\caption{Throughput performance of the three algorithms over 300 iterations. Aidos converges fastest, reaching 0.27 within 50 generations with only minor gains afterwards.}
\label{fig:throughputCI}
\end{figure}

\begin{figure}[t]
\centering  
\includegraphics[width=0.9\linewidth, clip, 
trim=0.1cm 0.1cm 0.1cm 0.1cm]
{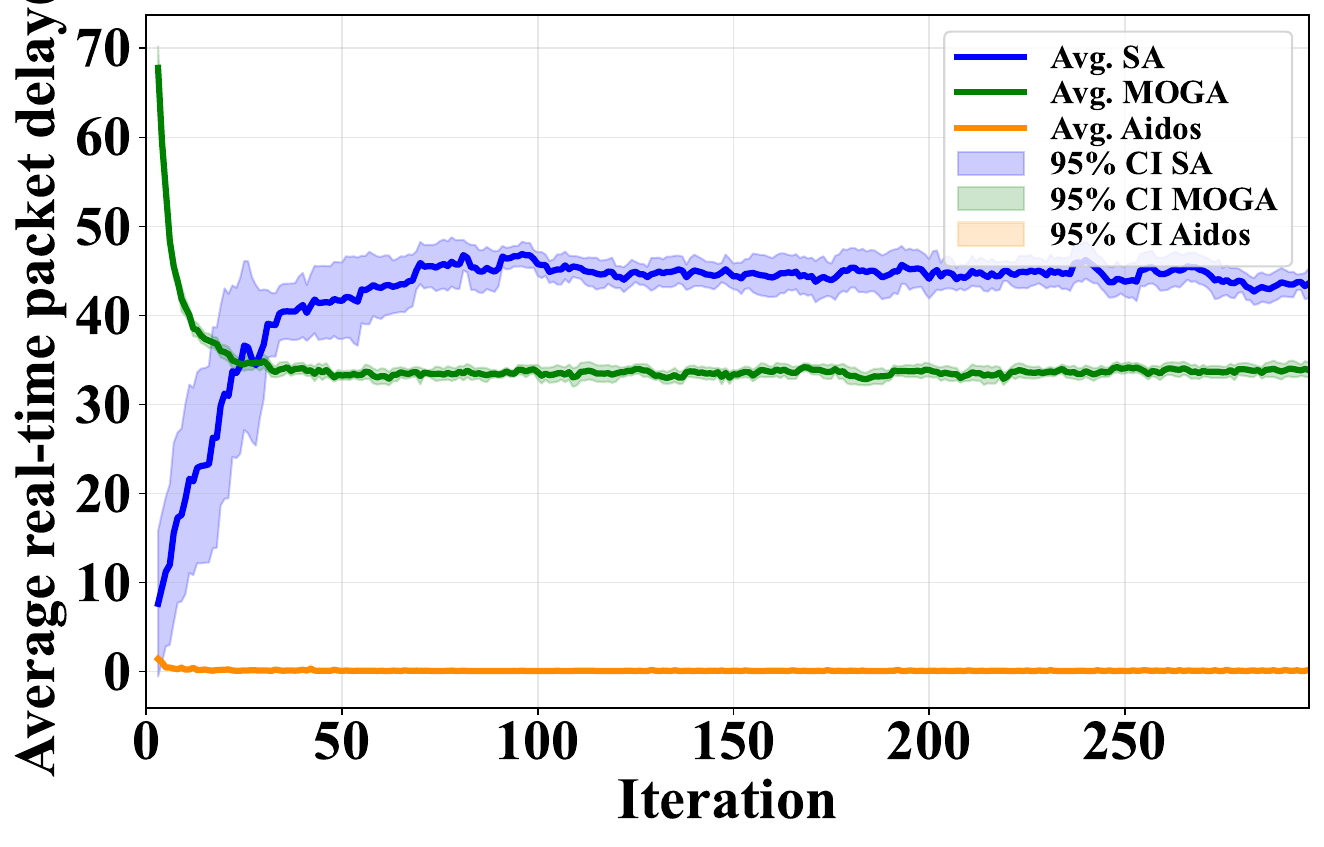} 
\caption{Delay performance of the three algorithms over 300 iterations. Aidos’s latency plunges early and settles near 0.5 s.}
\label{fig:delayCI}
\vspace{-0.3cm}
\end{figure}

\textit{Scalability:}
In next-generation LEO constellations such as Starlink and OneWeb, cells are significantly smaller, and a single satellite covers thousands of cells.
Scalability is therefore a key evaluation criterion.
Aidos is evaluated under the operational Starlink constellation, with the Beijing ground station (39.91° N, 116.40° E) designated as the observation point. 
By varying the minimum receive elevation angle at the ground station, we assess the scalability of Aidos across multiple scenarios.

\begin{figure*}[t]
    \centering
    \subfloat[Throughput in scenario 1\label{fig:144}]{%
        \includegraphics[width=0.32\textwidth]{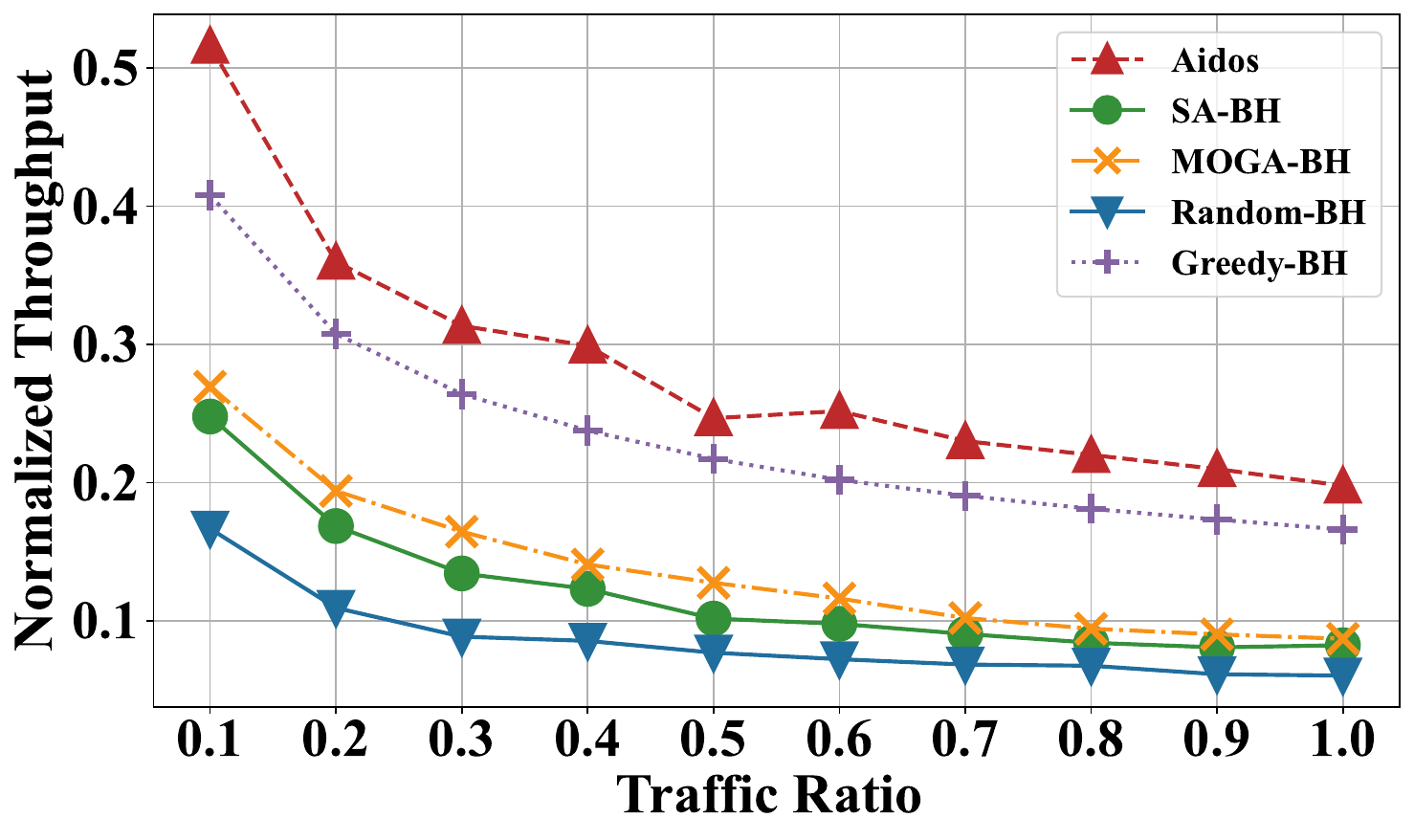}
    } \hfill
    \subfloat[Throughput in scenario 2\label{fig:240}]{%
        \includegraphics[width=0.32\textwidth]{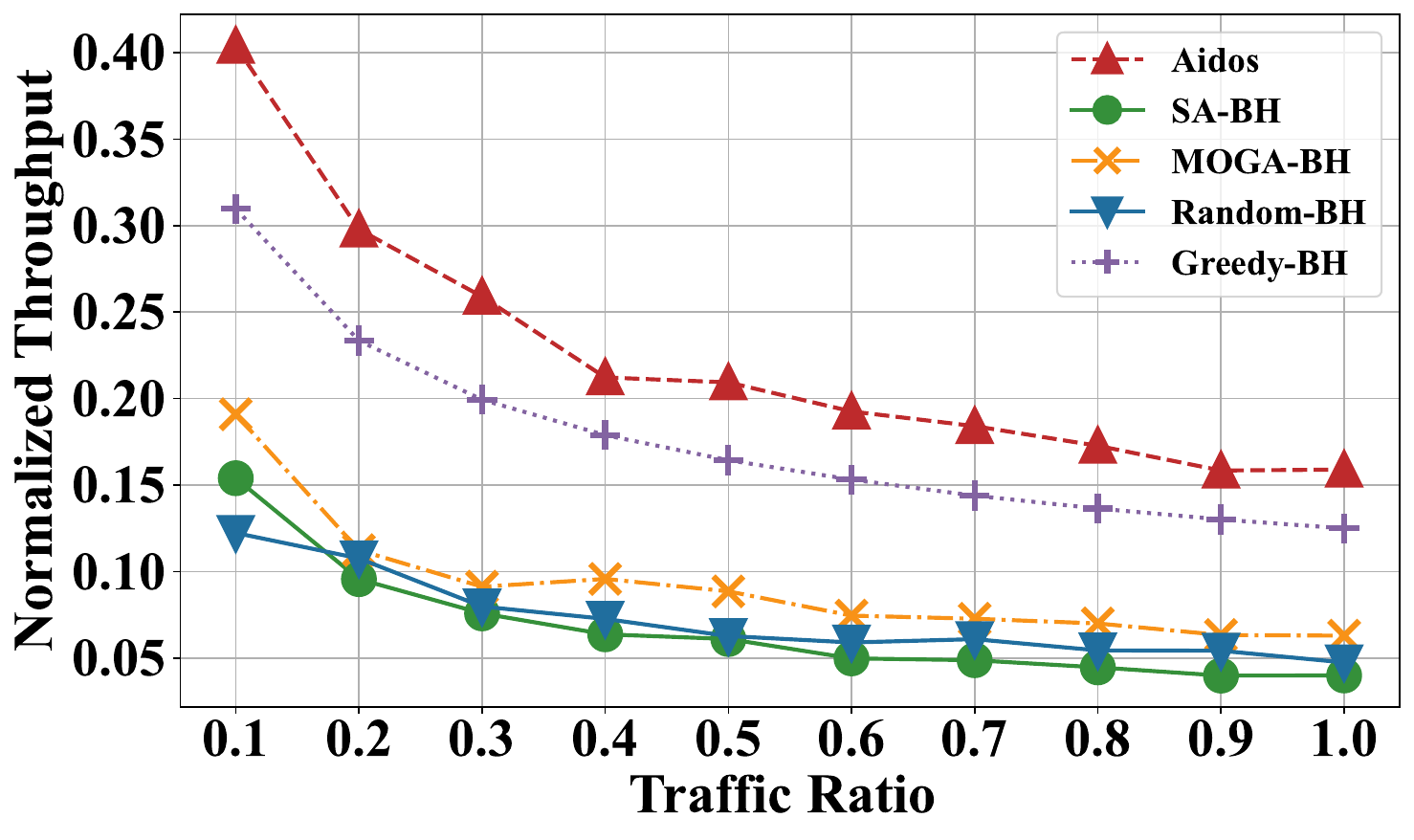}
    } \hfill
    \subfloat[Throughput in scenario 3\label{fig:336}]{%
        \includegraphics[width=0.32\textwidth]{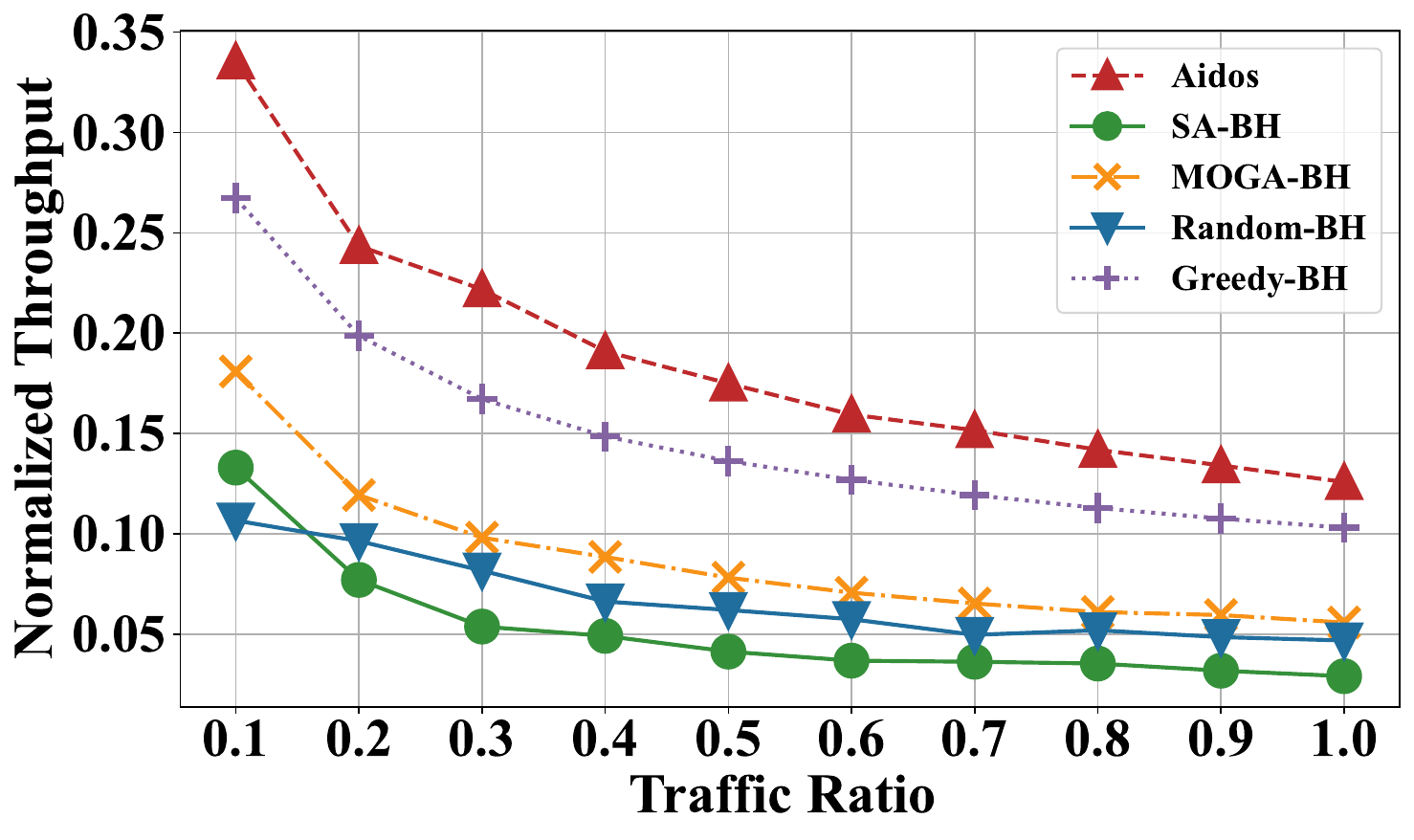}
    }
    \caption{Across the three enlarged scenarios, Aidos retains the highest aggregate throughput, outperforming MOGA-BH by 79.2\%–123.1\%. Both SA-BH and MOGA-BH show noticeable throughput degradation as the beam count rises, highlighting Aidos’s superior scalability.}
    \label{fig:large_throughput}
\end{figure*}

\begin{figure*}[t]
    \centering
    \subfloat[Latency in scenario 1\label{fig:144}]{%
        \includegraphics[width=0.32\textwidth]{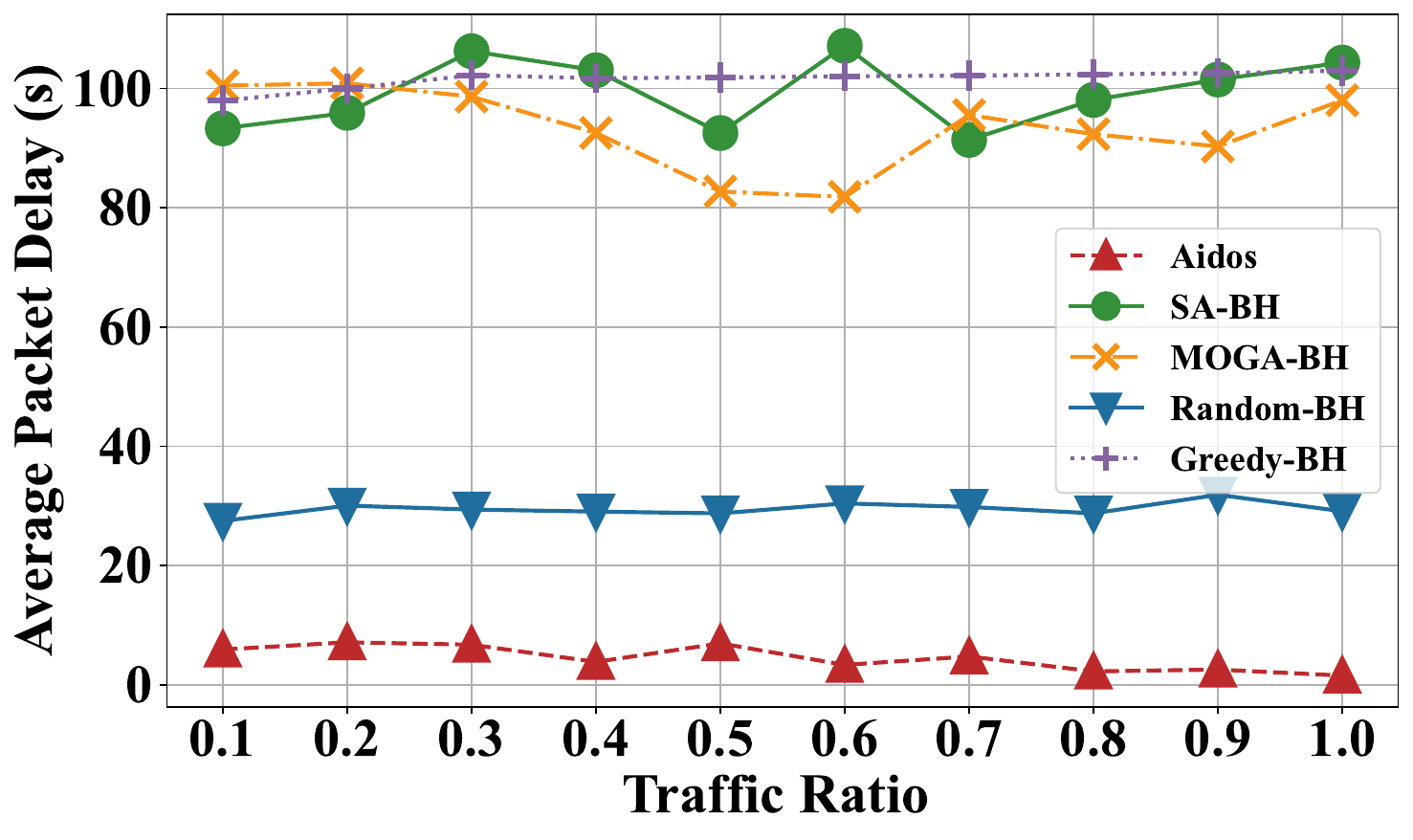}
    } \hfill
    \subfloat[Latency in scenario 2\label{fig:240}]{%
        \includegraphics[width=0.32\textwidth]{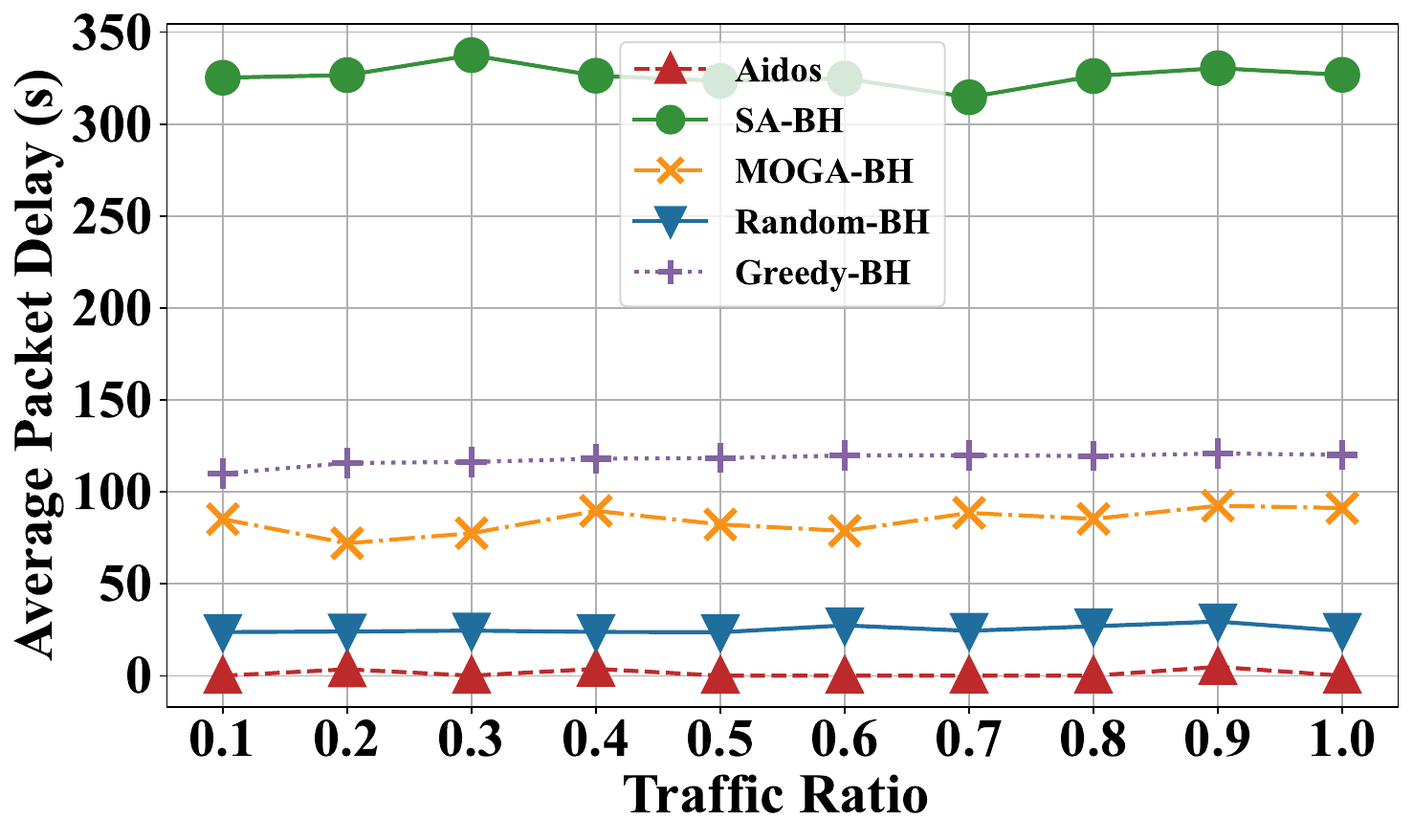}
    } \hfill
    \subfloat[Latency in scenario 3\label{fig:336}]{%
        \includegraphics[width=0.32\textwidth]{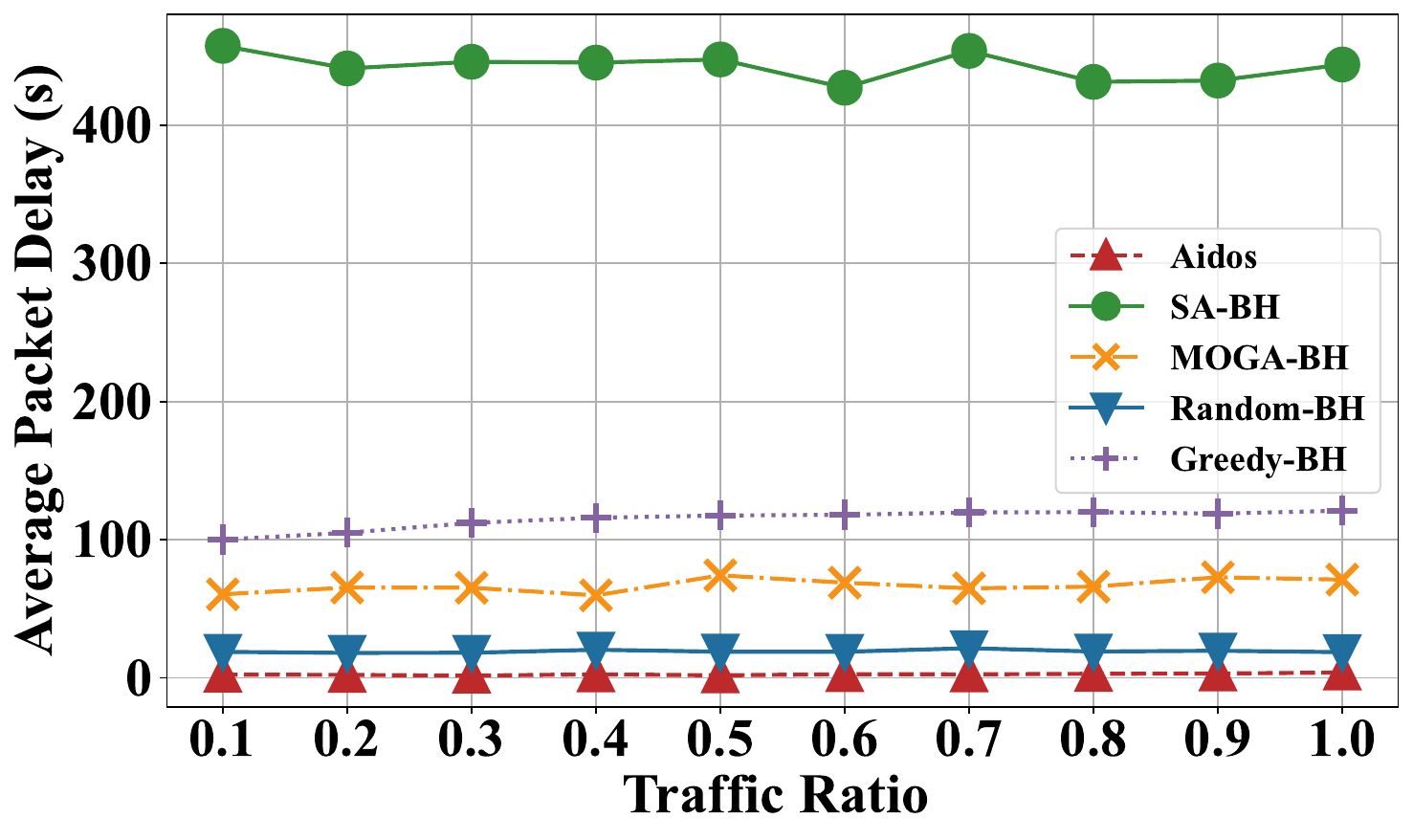}
    }
    \caption{Aidos keeps latency below 5 s over all beam counts, whereas SA-BH’s delay grows sharply and MOGA-BH exceeds real-time bounds under large-scale settings. The trend confirms that only Aidos scales efficiently in terms of latency.}
    \label{fig:large_delay}
\end{figure*}

\subsection{Overall Performance}

The experimental results report throughput and latency trends under different ground-traffic intensities.

As illustrated in Fig.~\ref{fig:100throughput}, Aidos exhibits a marked advantage in throughput. At a traffic ratio of 0.1, Aidos achieves a throughput of 0.559.
The gains over SA-BH, MOGA-BH, Random-BH, and Greedy-BH are about 93.5\%, 79.2\%, 232.1\%, and 33.9\%, respectively.
As the traffic ratio increases, all algorithms face reduced headroom and show a consistent decline.
Nevertheless, Aidos undergoes the smallest degradation.
Even under full load, where the traffic ratio equals 1, Aidos still exceeds the second-best method, Greedy-BH, by 26.67\%. By contrast, MADRL-BH performs the worst because the agents fail to learn an effective policy, confirming that the method us unsuitable for large-scale BH scenarios.

Fig.~\ref{fig:100delay} shows the latency results. 
Aidos maintains stable latency across traffic levels and remains below 70 ms.
The value meets the common voice target of less than 150 ms\cite{ITU-TG1051Amd1}.
Compared with SA-BH, MOGA-BH, Random-BH, and Greedy-BH, Aidos reduces average latency by about 99.45\% to 99.9\%.

\subsection{Ablation Study Results}

Fig.~\ref{fig:ablation} and ~\ref{fig:LR} 
reports ablation results for the four key modules in the Aidos and confirms their independent contributions.
As illustrated in Fig. 12(a), 
the TARK module is the most critical component.
It expands the searchable space and increases the HV by 40\%–60\%.
Fig. 12(b) and Fig. 12(c)
further reveal that 
STP-PMD and MSP both enhance population diversity and accelerate convergence; their HV gains are 12\%–25\% and 15\%–30\%, respectively.
In Fig. 13(a), 
the LR module refines already converged solutions. It improves latency by about 7\% without any throughput loss.
Finally, 
Fig.13(b) shows that local refinement also raises the overall HV by 5\%-10\%.

\subsection{Convergence Result}

Fig.~\ref{fig:throughputCI} and~\ref{fig:delayCI}
present throughput and latency performance over 300 iterations for Aidos, SA-BH, and MOGA-BH. As shown, Aidos exhibits the best convergence characteristics. 
In terms of throughput, Aidos reaches 0.27 within 50 generations and continues to improve slightly, whereas SA-BH and MOGA-BH require about 250 iterations to approach convergence. 
In terms of latency, Aidos exhibits a rapid decline during the initial iterations and stabilizes at approximately 100 ms. By contrast, MOGA-BH reduces to about 34\,s only after 40 generations, while SA-BH still fluctuates slowly at 45 s even after 100 generations.

Aidos reaches the convergence threshold after only 50 iterations:
the running mean varies by no more than 1\% over 20 consecutive generations.
MOGA-BH requires 150 generations and SA-BH still oscillates after 250 generations.

Aidos also shows the smallest standard-deviation change, highlighting its superior stability. Therefore, 300 iterations are sufficient to cover the steady-state phase for all three algorithms. 
With an early-stopping rule based on the target threshold, Aidos can cap the iteration limit at 50 generations and save more than 65\% of the computational budget.

\begin{figure}[t]
\vspace{-0.4cm}
    \centering
    \subfloat[$\theta_{\mathrm{elev}} \geq 50^\circ$]{
        \includegraphics[width=0.22\textwidth]{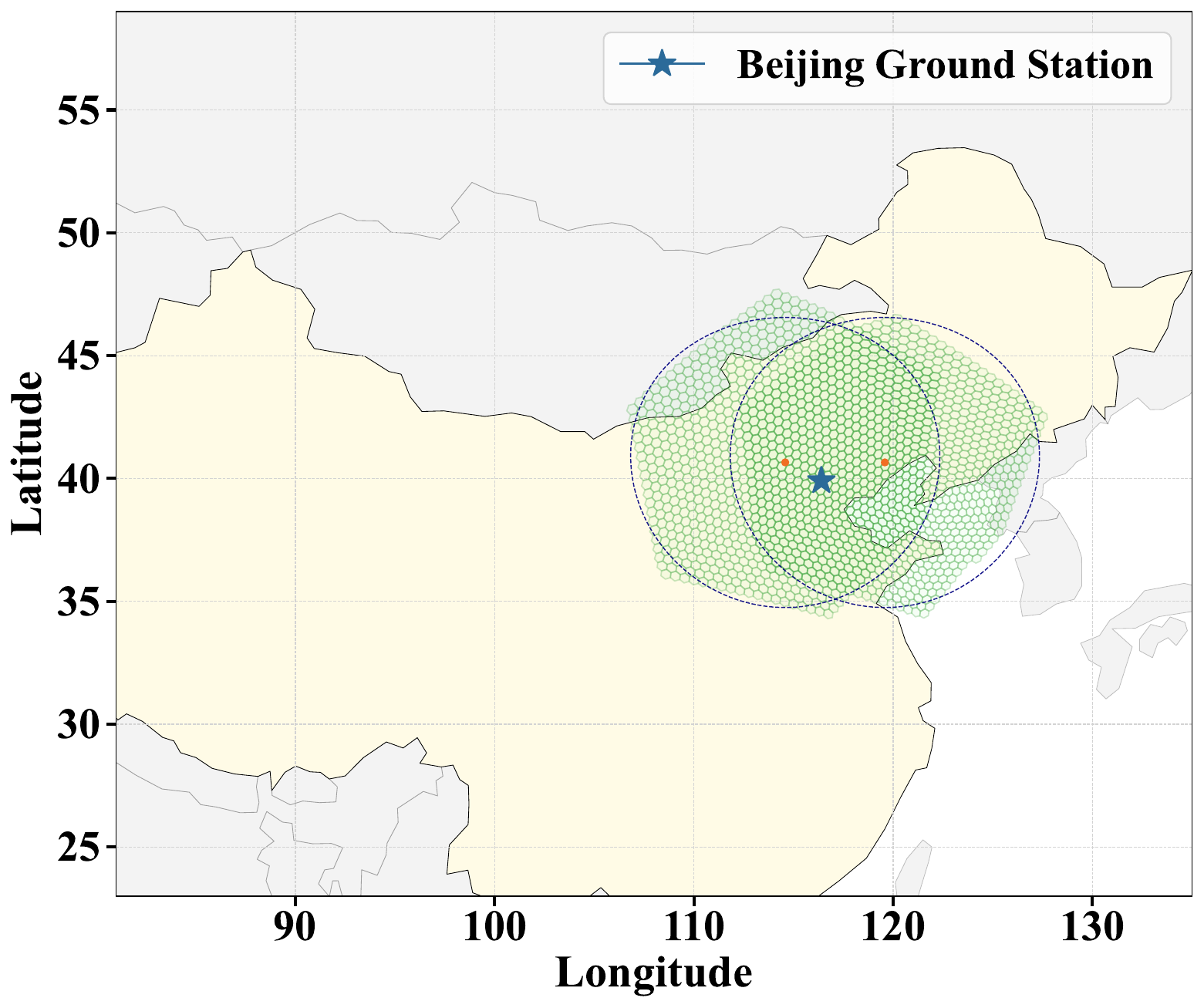}
    }
    \subfloat[$\theta_{\mathrm{elev}} \geq 42^\circ$]{
        \includegraphics[width=0.22\textwidth]{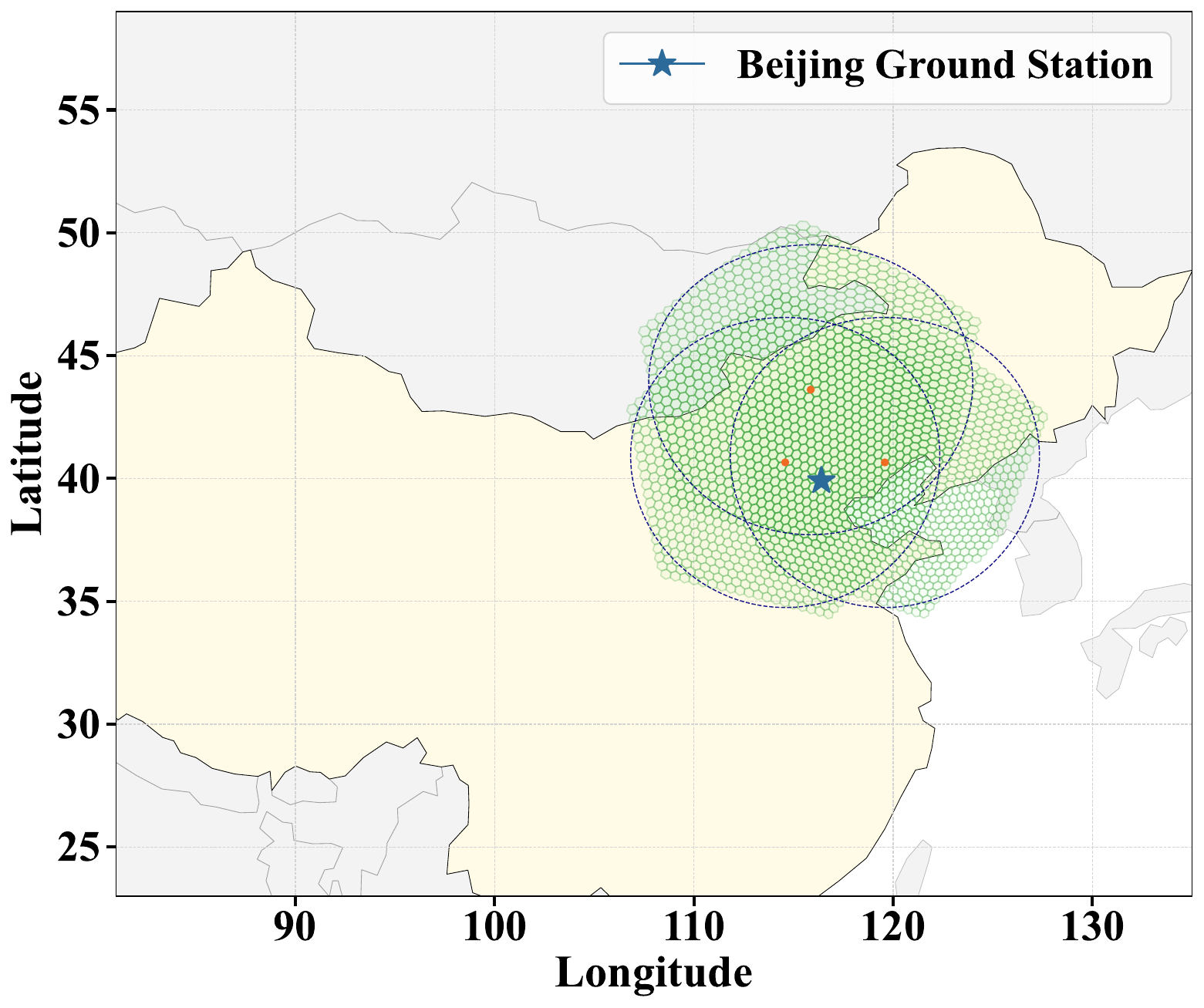}
    }\\[1ex]
    \subfloat[$\theta_{\mathrm{elev}} \geq 38^\circ$]{
        \includegraphics[width=0.22\textwidth]{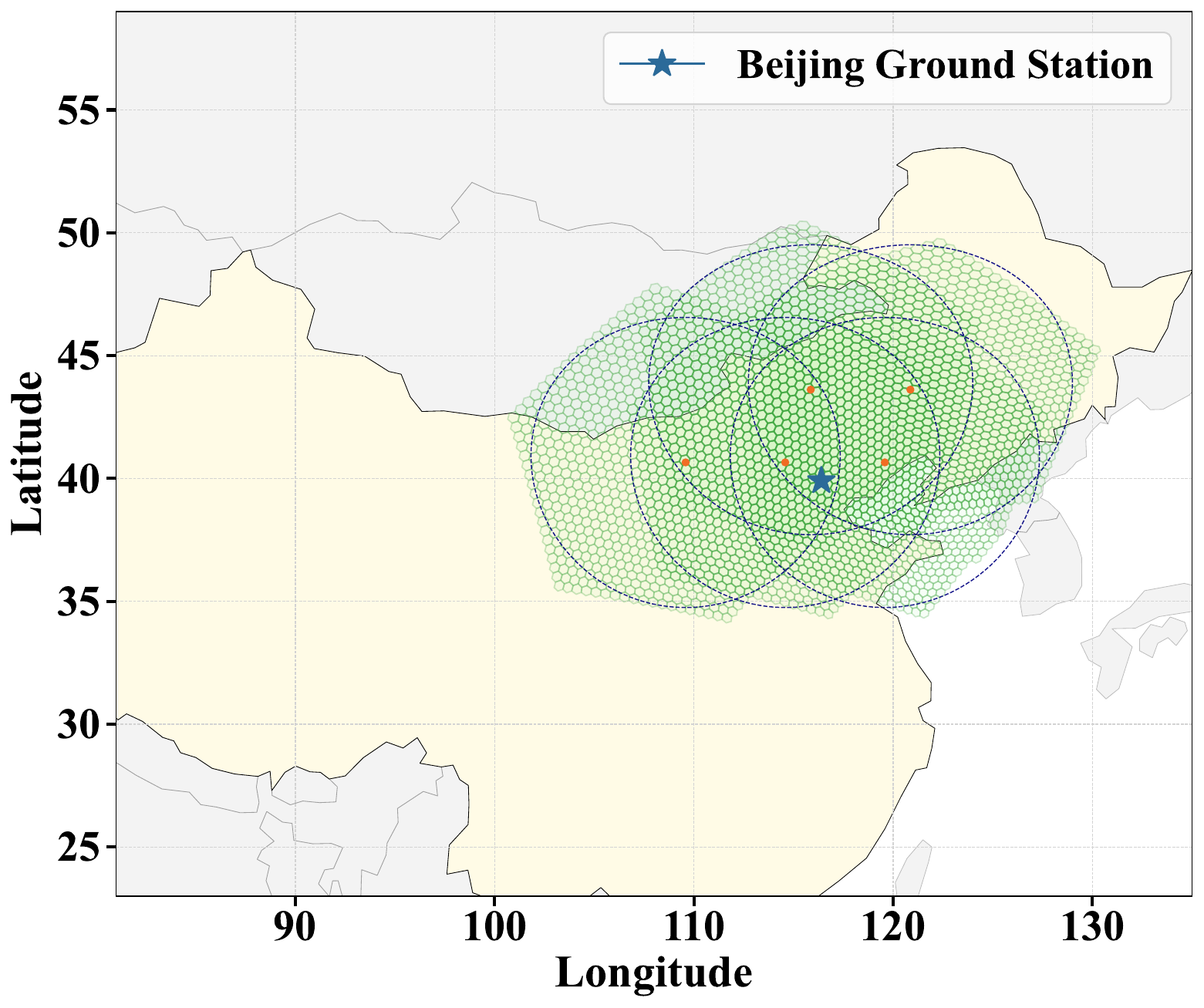}
    }
    \subfloat[$\theta_{\mathrm{elev}} \geq 33^\circ$]{
        \includegraphics[width=0.22\textwidth]{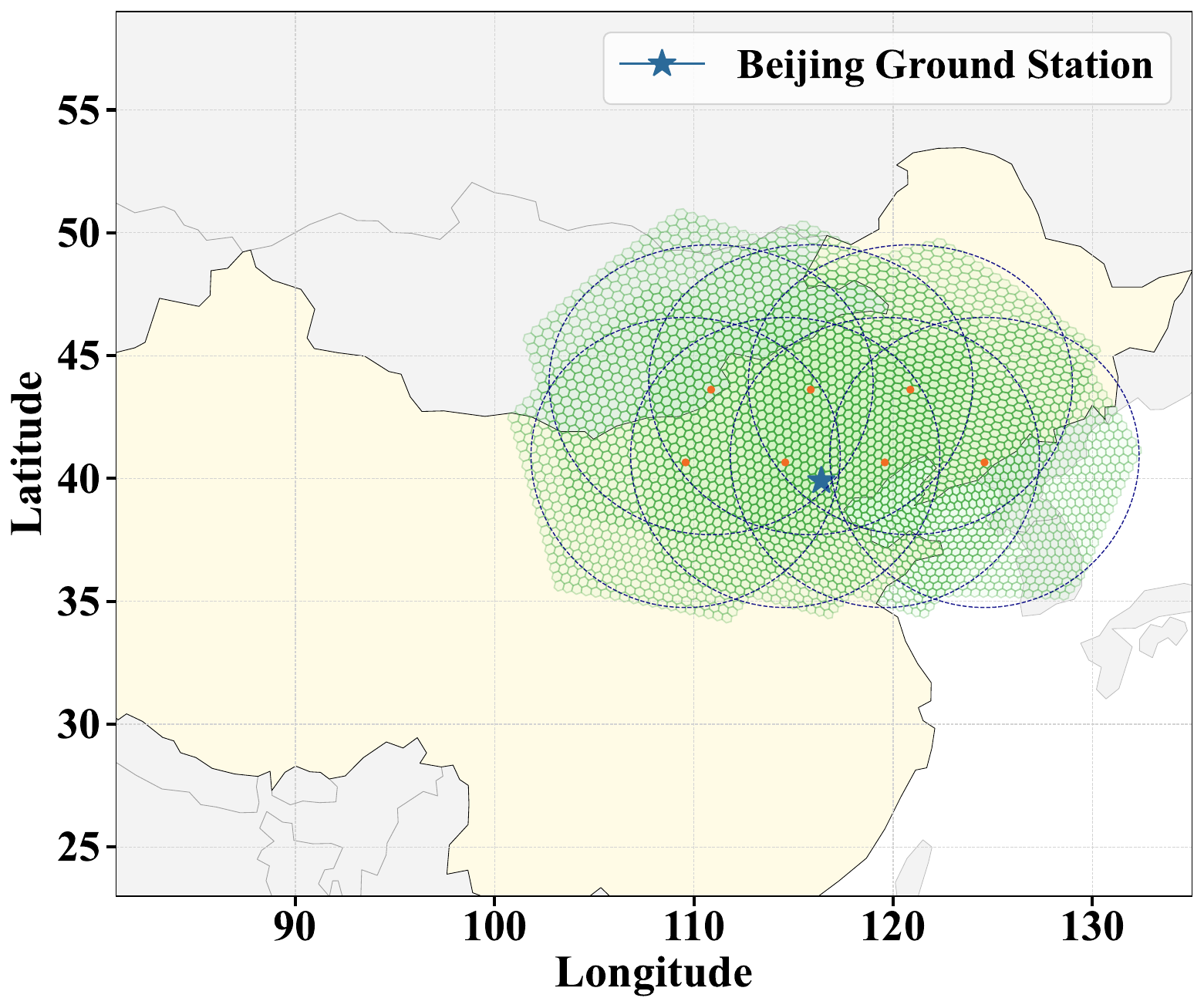}
    }
    \caption{Actual coverage of the Starlink constellation as observed from the Beijing ground station (39.91°N, 116.40°E), under minimum elevation angle thresholds of 50°, 42°, 38°, and 33°, respectively.}
    \label{fig:evel}
\vspace{-0.3cm}
\end{figure}

\begin{table}[t]
  \centering
  \renewcommand{\arraystretch}{1.4}  
  \setlength{\tabcolsep}{7.5pt}         
  \caption{Extended Scenario Parameters}
  \label{tab:scenario}
  \begin{tabular}{|c|c|c|c|c|}
    \hline
      & \makecell[c]{\textbf{Min.\ Elev.}}
      & \makecell[c]{\textbf{Visible}}
      & \makecell[c]{\textbf{Capacity}}
      & \makecell[c]{\textbf{Scenario Size}}\\
      & \makecell[c]{\textbf{Angle /$^\circ$}}
      & \makecell[c]{\textbf{Satellite}}
      & \makecell[c]{\textbf{/Gbps}}
      & \makecell[c]{\textbf{(cells, beams)}}\\
    \hline
    Scenario~0 & 50$^\circ$ & 2 & 19.2 & (1127, 96)\\ \hline
    Scenario~1 & 42$^\circ$ & 3 & 28.8 & (1667, 144)\\ \hline
    Scenario~2 & 38$^\circ$ & 5 & 48   & (2282, 240)\\ \hline
    Scenario~3 & 33$^\circ$ & 7 & 67.2 & (2693, 336)\\ \hline
  \end{tabular}
\end{table}

\subsection{Scalability Result}

Fig.~\ref{fig:evel} presents the actual Starlink coverage observed from the Beijing ground station (39.91°\,N, 116.40°\,E).
We set the minimum receive elevation to 50°, 42°, 38°, and 33°.
By counting the number of visible satellites and the total H3 cells within the coverage footprint, we obtain three extended experimental scenarios as listed in TABLE~\ref{tab:scenario}.

Fig.~\ref{fig:large_throughput} and~\ref{fig:large_delay} reports the throughput and latency in three extended scenarios. The results show that not all algorithms scale well.
Aidos exhibits the best scalability and handles large networks as the number of beams increases.
It maintains the highest throughput, exceeding MOGA-BH by 79.2\%–123.1\%, and keeps latency below 5 s.
In contrast, the performance of SA-BH degrades markedly under high beam counts.
Although MOGA-BH surpasses SA-BH, it still falls short of Aidos overall. 
Due to limits from evolutionary step size and computation time, MOGA-BH is better suited to offline planning where accuracy is prioritized and real-time demands are low.

\subsection{Computation Time}
TABLE~\ref{tab:avg_time} indicates that Aidos’s average runtime increases almost linearly with the problem size, indicating that instance scale and iteration count are the dominant factors in its computational cost.
With the iteration limit uniformly set to 150, Aidos still completes within 2\,min for the largest scenario.
The computation time is about 54.5\% lower than MOGA-BH.
SA-BH, Greedy-BH, and Random-BH each run in under 3 s and remain real time even at maximum scale, but their solution quality is markedly lower than Aidos.

Based on the convergence analysis, 
the iteration limit for Aidos can be reduced to 50 without sacrificing solution optimality.
With this cap, the average runtime in the largest scenario falls to 23 s.
These results indicate that the adopted outer-loop setting provides a practical balance between solution quality and runtime. Using a significantly larger iteration budget would bring only limited improvement while increasing the computation time.
In this scenario, the minimum elevation is 33°, and a 550 km LEO satellite has an overpass window of about 300 seconds.
Hence, Aidos can recompute about 13 times within one visibility window. 
This provides ample opportunity for online re-planning and ensuring strong real-time performance.

\begin{table}[t]
  \centering
  \setlength{\tabcolsep}{4pt}   
  \renewcommand{\arraystretch}{1.4} 
  \caption{Average Computation Time of Different Algorithms}
  \label{tab:avg_time}

  \begin{tabular}{|l|c|c|c|c|}
    \hline
    \multirow{2}{*}{\textbf{Algorithm}} &
      \multicolumn{4}{c|}{\textbf{Average Computation Time(s)}} \\ \cline{2-5}
      & \textbf{Scenario 0} & \textbf{Scenario 1} & \textbf{Scenario 2} & \textbf{Scenario 3} \\ \hline
    MOGA-BH                & 67.793 & 88.560 & 128.374 & 151.451 \\ \hline
    Aidos (150 iters)      & 30.903 & 49.050 & 69.7837 & 86.140  \\ \hline
    \textbf{Aidos (50 iters)} & \textbf{9.298} & \textbf{17.204} & \textbf{19.971} & \textbf{22.952} \\ \hline
  \end{tabular}
\end{table}

\section{Conclusion}

In this paper, we present Aidos, a hybrid optimization algorithm for computing the BHTP. 
The method integrates traffic-aware random-key encoding with a multi-objective metaheuristic search. It further adopts sliding-window Beta resampling within an adaptive distribution evolution to improve efficiency and solution quality.
Compared with MOGA-BH and MADRL-BH, Aidos increases throughput by 79.2\% and 247.2\%, respectively, and reduces latency by 99.45\%. 
Scalability experiments across multiple cell counts show that, with the iteration limit set to 50 generations, Aidos 
achieves an average runtime of 9.3 s in the 1,127-cell scenario.
The value fits within a 300 s satellite overpass window and enables multiple online re-optimizations.
These findings confirm that Aidos is well-suited for real-time BHTP synthesis in large-scale NGSO constellations.

Future work will investigate more scalable distributed implementations and multi-satellite cooperative scheduling architectures for LEO mega-constellation systems. It will also extend the current BH framework to incorporate resource allocation in an additional power dimension. This may improve allocation flexibility, but also results in a more challenging higher-dimensional coupled optimization problem.

\section*{Acknowledgment}
The research is sponsored by the Natural Science Foundation of Shanghai under Project No. 25ZR1402021.

\ifCLASSOPTIONcaptionsoff
  \newpage
\fi

\bibliographystyle{IEEEtran}  
\bibliography{main_bib}           

\vspace{-2.0em}


\begin{IEEEbiography}[{\includegraphics[width=1in,height=1.25in,clip,keepaspectratio]{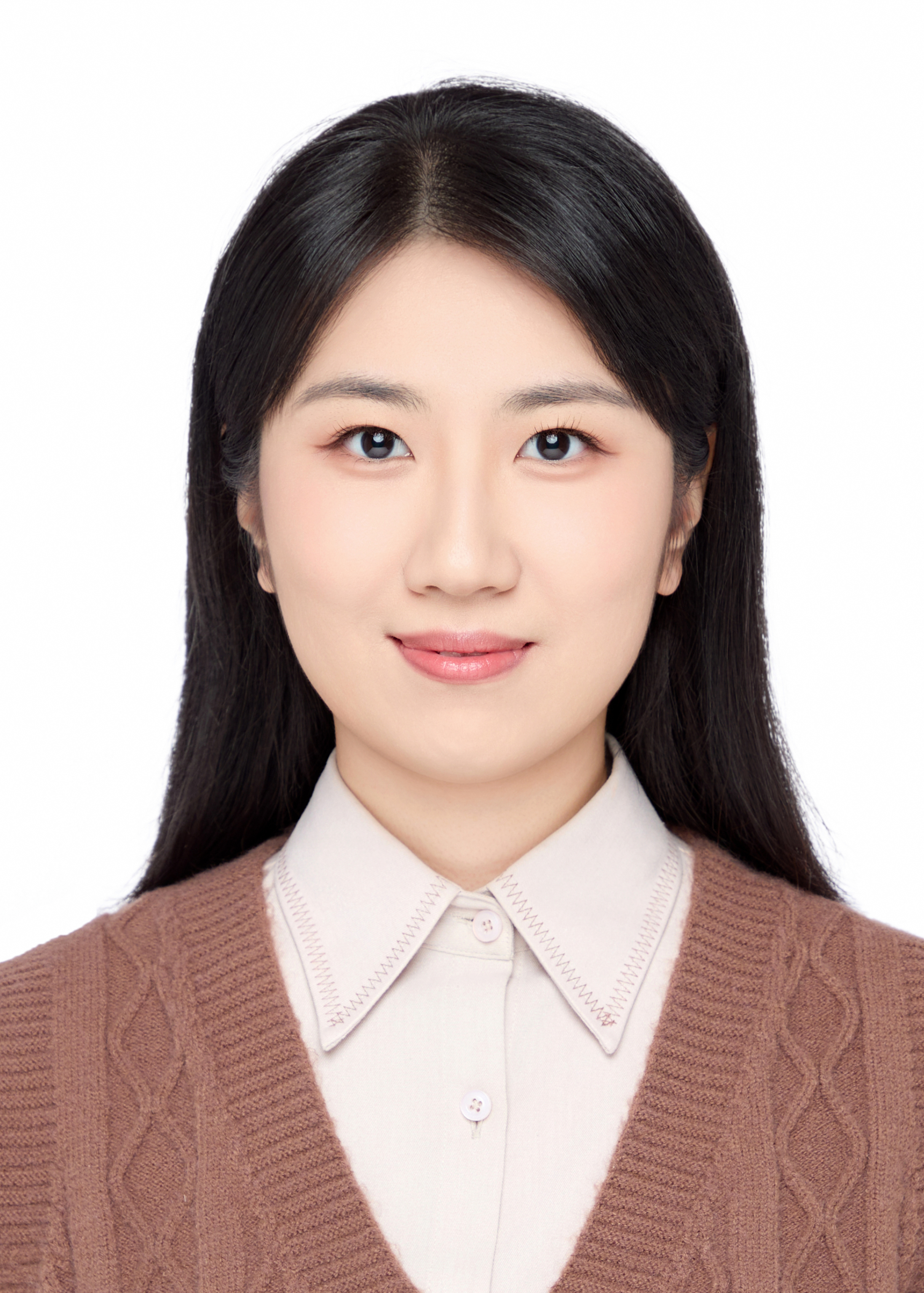}}]{Lingkai Zhao}
received the B.S. degree in Remote Sensing  and the M.S. degree in Communication Engineering , both from Harbin Institute of Technology, Harbin, China. She is currently pursuing the Ph.D. degree at Fudan University, Shanghai, China.
Her research interests include satellite communications and beam hopping.
\end{IEEEbiography}

\begin{IEEEbiography}[{\includegraphics[width=1in,height=1.25in,clip,keepaspectratio]{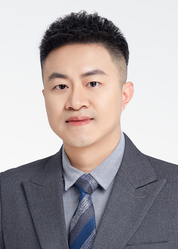}}]{Zhe Chen}
received his PhD in Computer Science from Fudan University, China, with a 2019 ACM SIGCOMM China Doctoral Dissertation Award. He is an Assistant Professor in the School of Computer Science at Fudan University and the Co-Founder of AIWiSe Ltd. Inc. Before joining Fudan University, he worked as a research fellow at NTU for three years, and his research achievements, along with his efforts in launching products based on them, have thus earned him 2021 ACM SIGMOBILE China Rising Star Award recently.
\end{IEEEbiography}

\begin{IEEEbiography}[{\includegraphics[width=1in,height=1.25in,clip,keepaspectratio]{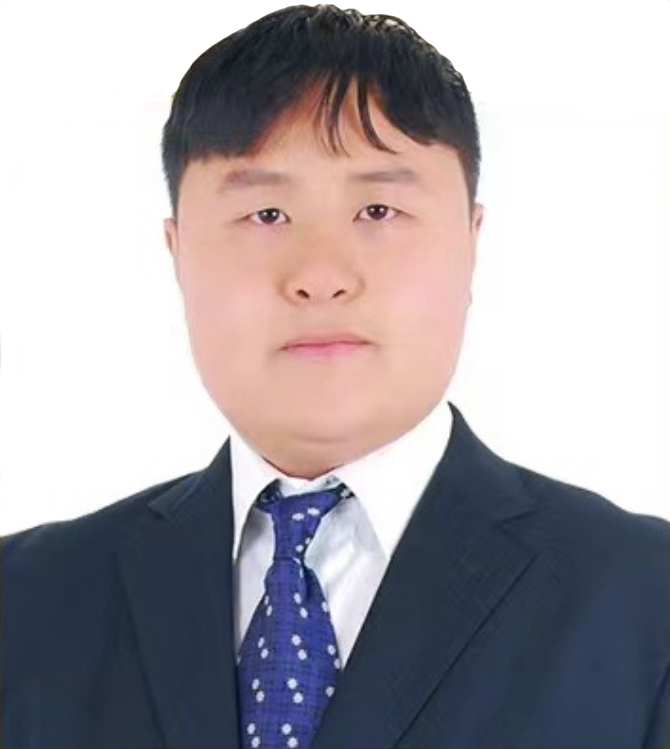}}]{Kun Qiu}
received his B.Sc. from Fudan University in 2013 and his PhD from Fudan University in 2019. He works for Intel as a software engineer from 2019 to 2023. He joined Fudan University in 2023 as an Assistant Professor in the School of Computer Science at Fudan University. His research interests include computer networks and computer architecture. He is a member of IEEE, ACM, and CCF.
\end{IEEEbiography}

\begin{IEEEbiography}[{\includegraphics[width=1in,height=1.25in,clip,keepaspectratio]{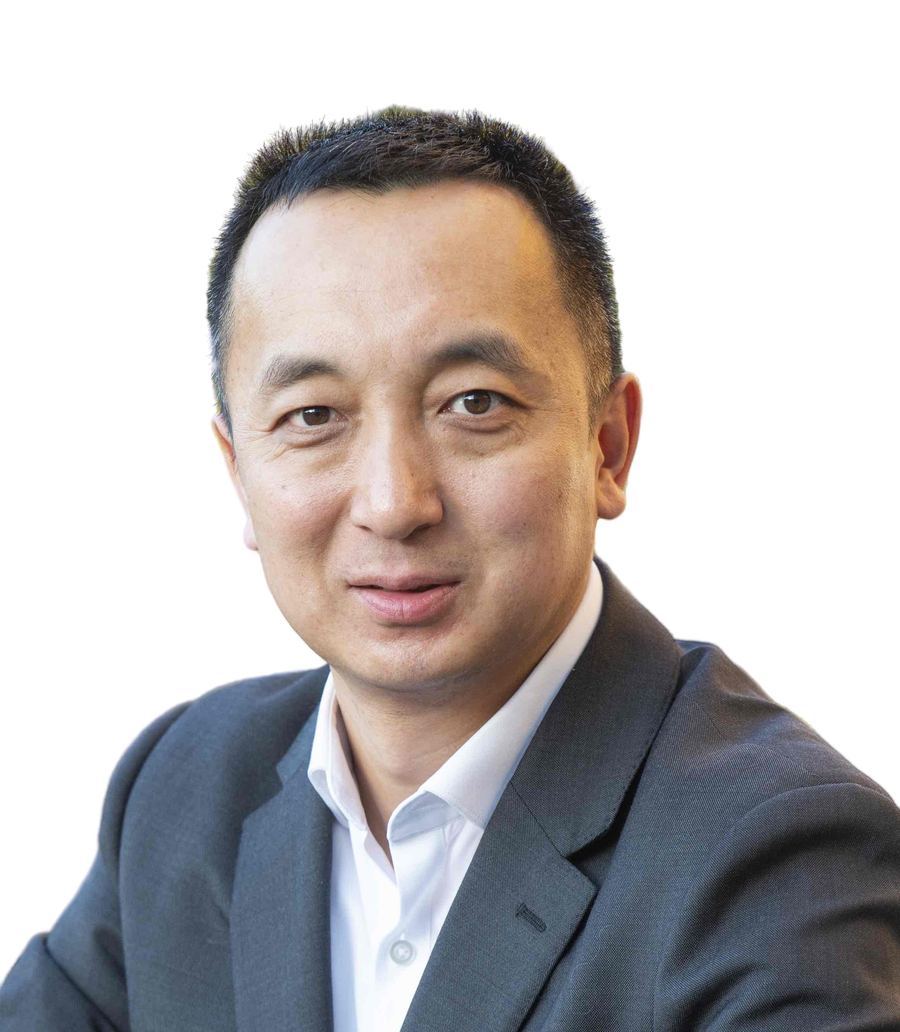}}]{Yue Gao}
received his PhD from the Queen Mary University of London, UK, in 2007. He is a Chair Professor at the School of Computer Science, Director of the Intelligent Networking and Computing Research Centre at Fudan University, China and a Visiting Professor at the University of Surrey, UK. His research interests include smart antennas, sparse signal processing and cognitive networks for mobile and satellite systems. He is a Fellow of the IEEE.
\end{IEEEbiography}

\end{document}